\documentclass[journal,draftclsnofoot,onecolumn,12pt]{IEEEtran}

\usepackage{amsthm,amssymb,graphicx,multirow,amsmath,color,amsfonts}
\usepackage[update,prepend]{epstopdf}
\usepackage[noadjust]{cite}
\usepackage[latin1]{inputenc}
\usepackage{tikz}
\usetikzlibrary{arrows,calc}		
\usepackage{bbm} 
\usepackage{pdfpages}
\usepackage{tabulary}
\usepackage{multirow}
\usepackage{comment}
\usepackage{graphicx}
\usepackage[font=small,labelfont=bf]{caption}
\usepackage{subcaption}

\def\nb0{{\mathbf{0}}}
\def\nb1{{\mathbf{1}}}

\newtheorem{lemma}{Lemma}

\newtheorem{definition}{Definition}

\newtheorem{theorem}{Theorem}

\newtheorem{remark}{Remark}

\allowdisplaybreaks 

\begin{document}
\graphicspath{{./Figures/}}
\title{
Ambient Backscatter Systems: Exact Average Bit Error Rate under Fading Channels
}
\author{
J. Kartheek Devineni and Harpreet S. Dhillon
\thanks{J. K. Devineni and H. S. Dhillon are with Wireless@VT, Department of ECE, Virginia Tech, Blacksburg, VA (email: kartheekdj@vt.edu and hdhillon@vt.edu). \hfill Manuscript updated: \today.} 
}

\maketitle

\begin{abstract}
The success of Internet-of-Things (IoT) paradigm relies on, among other things, developing energy-efficient communication techniques that can enable information exchange among billions of battery-operated IoT devices. With its technological capability of simultaneous information and energy transfer, ambient backscatter is quickly emerging as an appealing solution for this communication paradigm, especially for the links with low data rate requirement. In this paper, we study signal detection and characterize exact bit error rate for the ambient backscatter system. In particular, we formulate a binary hypothesis testing problem at the receiver and analyze system performance under three detection techniques: a) {\em mean threshold (MT)}, b) {\em maximum likelihood threshold (MLT)}, and c) {\em approximate MLT}. Motivated by the energy-constrained nature of IoT devices, we perform the above analyses for two receiver types: i) the ones that can accurately track channel state information (CSI), and ii) the ones that cannot. Two main features of the analysis that distinguish this work from the prior art are the characterization of the exact conditional density functions of the average received signal energy, and the characterization of exact average bit error rate (BER) for this setup. The key challenge lies in the handling of correlation between channel gains of two hypotheses for the derivation of joint probability distribution of magnitude squared channel gains that is needed for the BER analysis.
\end{abstract}

\begin{IEEEkeywords}
Ambient backscattering, Bit error rate (BER), Internet of Things, Hypothesis testing, Noncentral chi-squared distribution.
\end{IEEEkeywords}

\section{Introduction} \label{sec:intro}

Ambient backscatter, with its technological capability of enabling low-rate and low-power communication among energy-constrained devices, is considered as a promising solution for the reliable exchange of data in the Internet-of-Things (IoT) paradigm. The main premise of ambient backscatter is to use omnipresent ambient electromagnetic (EM) waves, such as the radio frequency (RF) waves, cellular/WiFi or television (TV) signals, to both harvest energy at small IoT devices as well as to use these existing waves as carriers for data transmission. The utilization of backscattering mechanism for data modulation precludes the requirement of power-intensive RF-chain components like RF mixers, analog-to-digital converters (ADCs) and digital-to-analog converters (DACs), which greatly reduces the energy requirements of the circuit~\cite{shyam13,shyam16}. Such a technology is especially attractive for IoT devices deployed at hard-to-reach locations for which recharging or replacing batteries may not be economically viable. The ubiquitous presence of wireless networks provide a reliable source of EM waves that can be utilized by such devices for data transmission using backscattering. Due to the wide-ranging potential advantages of this technology, it is of immediate interest to characterize various aspects of its performance accurately. In this paper, we provide an exact characterization of BER for these ambient backscatter systems in a flat fading channel both in the presence and absence of CSI.

\subsection{Related Work}  \label{sec:related}

Although ambient backscatter communications have gained prominence recently, initial research on the fundamentals of backscatter systems dates back to 1948 when it was first applied in radar systems \cite{stock1948}. Later in the early 1990s and 2000s, it found a prominent application in inventory tracking and identification through radio frequency identification (RFID) systems. A serious drawback of these systems compared to traditional point-to-point communications is the {\em two-way} propagation loss resulting in a limited communication range. This motivated the study of channel characteristics and distance limitations of the conventional backscatter systems in~\cite{gridur2009, smith2003}. To overcome this limitation, approaches such as bistatic backscatter \cite{sahalos2014} were explored for improving range. The use of coding techniques and multiple antennas for performance improvements was explored in \cite{roym2014, gridur2008, gridur2010,leigh2008}. The security and protocol aspects of backscatter systems to achieve reliable communication were investigated in \cite{peeters2014,saad2014}. 

A major drawback of the conventional backscattering systems is the need for a standalone equipment to send the source RF signals, which are scattered back by devices such as a moving vehicle or miniature tags. Ambient backscattering \cite{shyam13,shyam16} is the first successful implementation of backscatter systems that circumvents the need for extra hardware, thereby reducing the cost of infrastructure and maintenance. Some of the recent prototype implementations of the ambient backscatter include low-power communication to nearby devices by leveraging the TV/cellular waves \cite{shyam13}, multiple antenna and coding techniques for improved throughput and range, respectively \cite{shyam14}, passive Wi-Fi transmissions with very low circuit operational power \cite{shyam2014}, low-power self-interference cancellation techniques for full-duplex transmissions and frequency-modulation (FM) backscattering for smart and connected cities \cite{liu14}, inter-technology backscatter to convert Wi-Fi packets into bluetooth transmissions \cite{shyam2016}, and long range (LoRa) low-power communications in the battery-less devices \cite{vamsi17, varshney17}. These proof-of-concept systems have demonstrated the feasibility of practical implementation of the ambient backscatter technology.

On the other hand, investigation into the theoretical aspects of ambient backscatter like throughput, error rates, and performance is still in the nascent stage. Several important steps in this direction had been taken in \cite{Wang15,chintha15,chintha16, gao16, gao17,hu2015, wang17, chintha16vtc,yang17}.
The design of maximum-likelihood and equiprobable-error detectors was first investigated in \cite{Wang15}. The detection using non-coherent and semi-coherent techniques at a receiver without channel state information was studied in \cite{chintha15, chintha16, gao16, gao17}. The detection of ambient backscatter signal with multiple receive antennas was performed in \cite{hu2015}. The statistical-covariance based signal detection to improve the BER of the system was investigated in \cite{chintha16vtc}. The BER analysis of detection over ambient orthogonal frequency division multiplexing (OFDM) signals using interference cancellation techniques was investigated in \cite{yang17}.  The capacity and throughput limits of an ambient scattering system were studied in \cite{verde2016}. In \cite{kaibin17}, the performance analysis of ambient backscatter in a network setup was performed in terms of the coverage probability and the transmission capacity using stochastic geometry framework. 

The key enabler of the analysis in \cite{chintha15,Wang15,hu2015, chintha16, wang17, gao16, gao17,chintha16vtc,yang17} was the approximation of the probability density function (PDF) of average energy of the received signal as Gaussian distributed. Despite the progress made in detection and BER analysis of the ambient backscatter systems in the aforementioned works, the following two fundamental problems are still open: (i) the characterization of the exact distribution of average signal energy and (ii) the characterization of exact average BER in fading channels. Tackling these two problems is the main focus of this paper. Further details on the main contributions of the paper are provided next.

\subsection{Contributions and Outcomes} \label{sec:contributions}
  
\paragraph*{Exact conditional distributions and detection mechanisms}

We investigate signal detection in ambient backscatter for two types of receivers, which we refer to as: i) receiver with CSI ($\mathcal{R}_1$) and ii) receiver without CSI ($\mathcal{R}_2$). We show that the exact conditional density functions of the average energy of the received signal follow noncentral chi-squared distribution (NC-$\chi^2$). Characterization of the exact conditional signal distribution is an important component in the exact performance analysis, which differentiates our work from the earlier works that approximated this distribution as Gaussian \cite{Wang15, chintha15, chintha16, hu2015}. A binary hypothesis testing problem is formulated and the detection is performed by comparing the average energy of the signal to a threshold. Three detection strategies are considered for receiver $\mathcal{R}_1$: i) \emph{mean threshold (MT) detection} in which the threshold is calculated as the mean of conditional expectations of the average signal energies received under different hypotheses, ii) \emph{maximum likelihood threshold (MLT)  detection} in which the threshold is evaluated as intersection point of the exact conditional PDFs, and iii) \emph{Approximate MLT detection} where threshold is evaluated as the intersection point of approximations of the conditional PDFs. For receiver $\mathcal{R}_2$, differential encoding strategy is used at the transmitter to overcome the ambiguity in decoding process \cite{shyam13}. Simple threshold evaluation strategies, such as the MT threshold, are used in $\mathcal{R}_2$ because of the lack of complete channel information at the receiver in this case. 

\paragraph*{Joint distribution of correlated fading components}

A key challenge in the error analysis is the need to characterize the joint distribution of correlated fading components belonging to the different hypotheses. In particular, although the individual links in the system may experience independent fading, overlapping backscatter data onto radio signals eventually results in different but correlated fading components for the two hypotheses. 
A key driver of this evaluation is the independence of the fading component of alternate hypothesis conditioned on the fading component of null hypothesis. Further, characterization of the conditional BER in terms of the generalized Marcum Q-function allows us to come up with several system insights, which are discussed next.

\paragraph*{Insights}
 Using the conditional BER expressions, we deduce that the optimal performance of ambient backscatter is dependent only on SNR of the ambient signal and not on the individual strengths of the ambient signal and noise. This trend is similar to the performance of the standard binary phase shift keying (BPSK) modulation in the classical setup. Second, the decay rate of BER defined as the rate of depreciation is observed to decrease with the increasing
sample length $N$. This is in contrast to the constant BER decay rate observed when plotted against SNR of the signal. Third, the SNR gain of the system follows diminishing returns with increasing value of the sample length $N$. Further, our results show that there is no noticeable difference in the BER performance of the three detection threshold techniques considered in this work. Therefore, simpler techniques, such as the MT technique, can be implemented without much degradation of the system performance.  

\section{System Model} \label{sec:SysMod}

\subsection{System Setup and Backscatter Operation}  \label{sec:BSop}

We consider a pair of devices, of which one is a backscatter transmitter (BTx) and the other is a receiver (Rx). We assume the presence of modulated carrier waves generated by a source in the environment, henceforth referred to as \emph{ambient waves} and \emph{ambient source} respectively, and the devices communicate through scattering of the incident ambient waves as described shortly. This is a valid assumption since such sources of carrier waves, for example TV, cellular or Wi-Fi networks, are almost omnipresent. Backscatter derives its name from the mode of information exchange, which is to communicate data through reflection of RF waves, and the procedure of backscattering ambient RF waves is called \emph{ambient backscatter}.  The word {\em backscatter} simply refers to the process of backward reflection of incident waves at a surface in different directions (called \emph{diffuse reflection}), unlike the typical single reflection observed at the surface of a mirror (called \emph{specular reflection}). This phenomenon is similar to how visible light is reflected by normal objects in all directions (not just a single reflection as in the case of a mirror) and is illustrated in Fig. \ref{fig:bcksctr}. In order to understand the operation of data modulation using backscatter, it is essential to look at the propagation of an EM wave between different surfaces. When EM waves propagating through free space hit the antenna, part of the wave is reflected back into free space due to the difference between the impedance of free space and antenna. The reflection coefficient $\alpha$ of the antenna, defined as the ratio of the amplitudes of the reflected wave to the incident wave, is given by:
\begin{align}
 \alpha &= \dfrac{A^-}{A^+} = \dfrac{\frac{Z_L}{Z_0} -1}{\frac{Z_L}{Z_0} +1} ,
\end{align}
where $Z_L$ is the impedance of the antenna and $Z_0$ is the impedance of free space. When $Z_L = Z_0$, the wave is completely absorbed with no reflection and the impedance matching is known as \emph{reflection-less} matching. On the other hand, for $Z_L = 0$ the wave is completely reflected. Therefore, one can simply change the impedance of the antenna according to the data to be transmitted to generate a modulated reflected wave.

This phenomenon is exploited by the backscatter systems in a slightly modified way, where data modulation on the reflected wave is realized by manipulating the impedance mismatch between antenna and the load component (which forms the main circuit). The main reason is that, in a typical backscatter device, the chip is directly placed at the terminals of the antenna \cite{seshu2005}. The load impedance is typically a complex value, due to which the wave reflection needs to be analyzed in terms of power \cite{kurokawa1965}. Hence, the reflection coefficient $\alpha$ at the boundary between antenna and load is characterized in terms of power rather than voltage. The reflection coefficient here, termed as \emph{power wave reflection coefficient}, is given by \cite{kurokawa1965}:
\begin{align}
 \alpha &= \dfrac{\frac{Z_L}{Z_a^*} -1}{\frac{Z_L}{Z_a} +1} ,
\end{align}
where $Z_L$ and $Z_a$ are the impedances of the load and antenna respectively and the symbol $*$ represents complex conjugate. In order to transfer all the power to load, the load impedance is set to $Z_L = Z_a^*$ which is known as \emph{maximum power transfer} matching. On the other hand, in order to reflect all the power, the load impedance is set to $Z_L=0$. Therefore, $Z_L = Z_a^*$ and $Z_L = 0$ are known as \emph{non-reflecting} and \emph{reflecting} states, respectively. The backscatter system can leverage  this to modulate data by tuning impedance of the load to vary reflection coefficient at this boundary. A simple modulation scheme is to tune the circuit between reflecting and non-reflecting states when transmitting bits $1$ and $0$, respectively. The system model for the ambient backscatter is illustrated in Fig. \ref{fig:sys_mod}. The devices in the network are assumed to either have their own power source or generate enough power from the ambient waves to run their circuits. The latter assumption is quite reasonable because the ambient backscatter systems are designed to operate at a very low power, of the order of few micro-watts.

\subsection{Channel Model}  \label{sec:ChMod}

\begin{figure}
    \centering
    \begin{minipage}[b]{0.50\columnwidth}
        \centering
        \includegraphics [width=\linewidth]{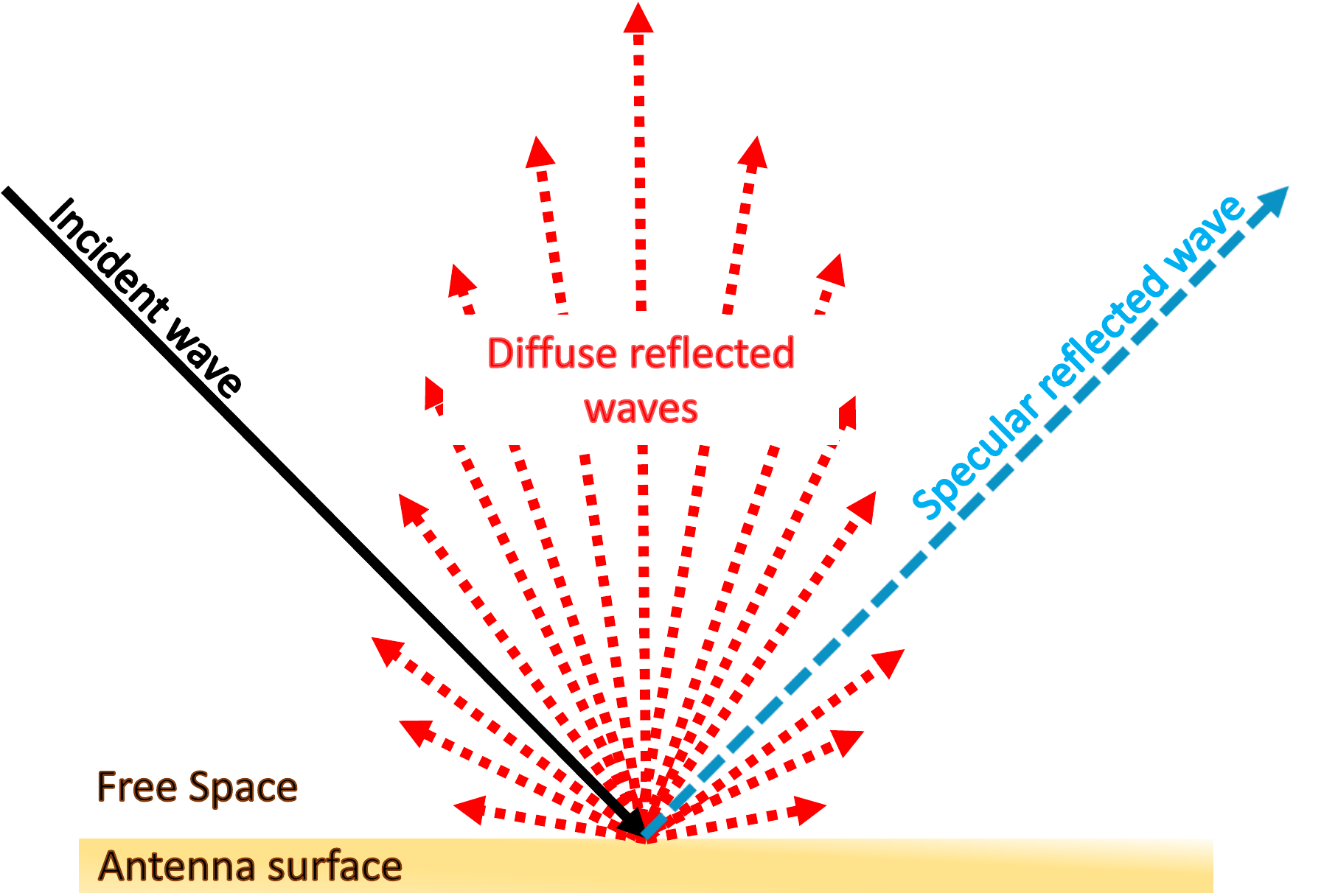}
        \caption{Illustration of diffuse reflection and specular reflection.}\label{fig:bcksctr}
    \end{minipage}
\hfill
    \begin{minipage}[b]{0.40\columnwidth}
        \centering
        \includegraphics [width=\linewidth]{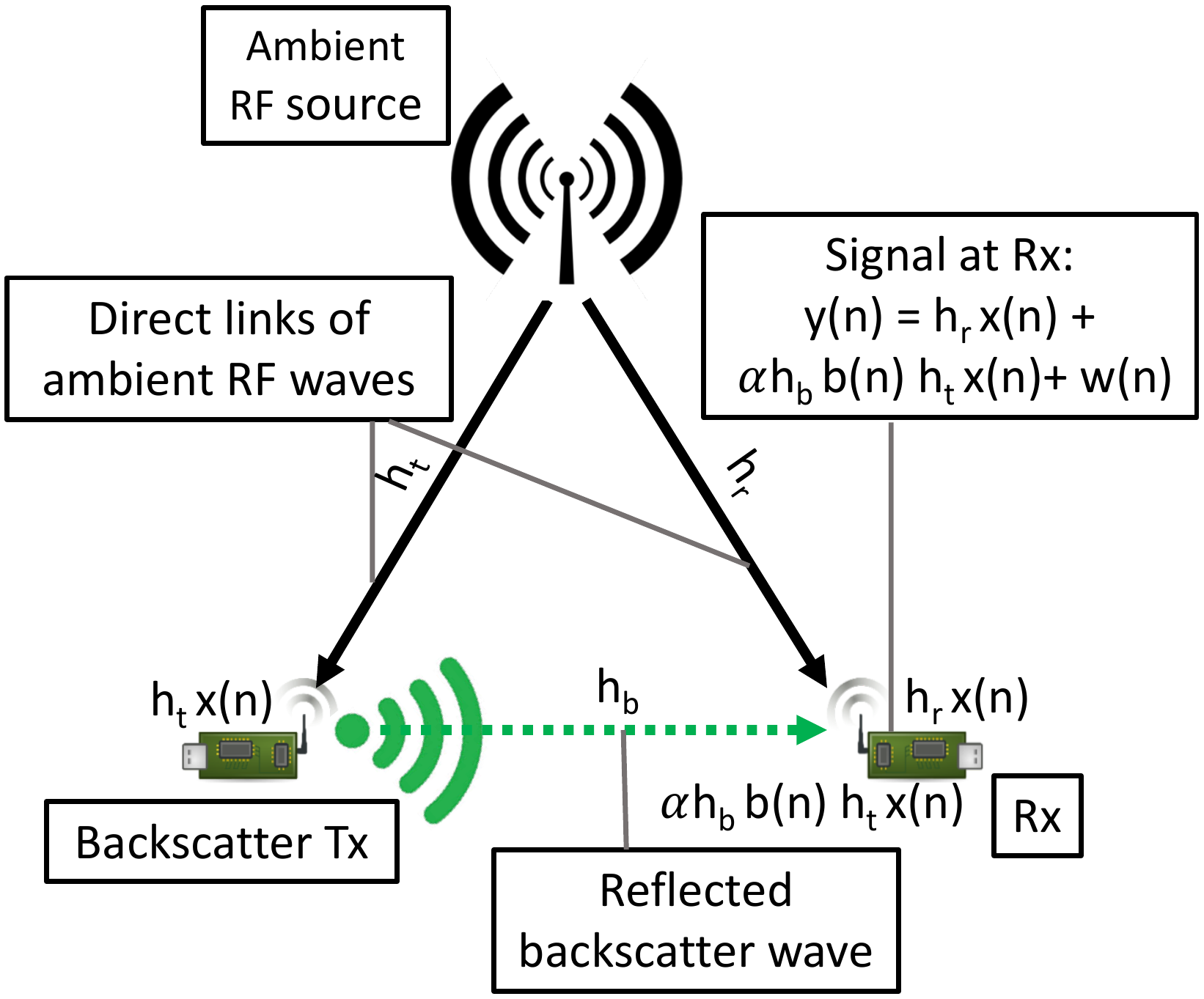}
	\caption{ System model of ambient backscatter communication system.}\label{fig:sys_mod}
    \end{minipage}
\end{figure}

In this paper we focus on flat Rayleigh fading channel. Handling more general fading distributions is a useful direction of future work. In the backscatter setup illustrated in Fig. \ref{fig:sys_mod}, there are two direct communication links, one each from \emph{ambient source} to transmitter and receiver, and one backscatter communication link, from transmitter to receiver. The fading components of the direct links to receiver and transmitter, and the backscatter link are independent, identically distributed and are denoted by $h_r$, $h_t$ and $h_b$, respectively. The average energy of the ambient signal is assumed to be unity and the variance $\sigma^2$ of zero mean additive complex Gaussian noise is varied to obtain different SNR values. For this reason, the exact units of signal energy are not needed and SNR is used as a measure of the signal strength in the distribution plots.
   
\subsection{Signal Model}  \label{sec:SigMod}

At the BTx, a simple binary on-off modulation scheme is implemented using reflecting and non-reflecting states to transmit digital bits. The desired signal at the Rx (shown in Fig. \ref{fig:sys_mod}) is the sum of two components, one directly received from the \emph{ambient source} and the other reflected from the BTx. The received signal of an ambient backscatter system is mathematically expressed as follows:
\begin{align}
	y(n) &= \underbrace{h_r x(n)}_{\text{radio signal}}+\underbrace{\alpha h_{b} \text{ } b(n) \text{ } h_t x(n)}_{\text{backscatter signal}} + \underbrace{w(n)}_{\text{i.i.d Gaussian noise}}, \label{eq:actmod}
\end{align}
where $x(n)$ and $w(n)$ are complex baseband ambient radio and additive complex Gaussian noise signals respectively, $b(n) \in \{0,1\}$ is the backscatter data and $\alpha$ is the reflection coefficient of the transmitter node at the boundary of antenna and circuit.

Assuming the data rate of backscatter communication is significantly lower than that of the data source (reasonable assumption for most IoT applications), the receiver can filter out the ambient source data $x(n)$ by simply averaging the energy of the received signal over $N$ samples, where $N$ is the window over which the backscatter data $b(n)$ remains constant~\cite{shyam13}. The average energy of $x(n)$ over the sample length $N$ is assumed to be a constant given by:
\begin{align}
\bar{E} = \frac {1} {N} \sum\limits_{n=1}^{N} |x(n)|^2 \label{eq: AvgE}.
\end{align}
By taking $b(n) =b$ over sample length $N$, the model in~\eqref{eq:actmod} can be simplified, as follows:
\begin{align}
	y(n) &= (h_r +  \alpha h_{b} h_t b) \, x(n) + w(n), & (1 \leq n \leq N).
\end{align}
To further simplify the model, received signal $y(n)$ can be expressed separately for each value of bit $b$ with the following fading components:
\begin{align}
	y(n) &= \begin{cases}
	h_0\, x(n) + w(n), & b = 0,\\
	h_1\, x(n) + w(n), & b= 1,
	\end{cases} \label{eq: BinHyp}
\end{align}
where $h_0 = h_r$ and $h_1 = h_r + \alpha h_b  h_t$ are fading components dependent on backscatter data $b$. The magnitude square of the fading components are denoted by $\mu = |h_0|^2$ and $\nu = |h_1|^2$.

\begin{remark}\label{rem: AMBbpsk}
	It should be noted that the fading terms $h_0$ and $h_1$ (also $\mu$ and $\nu$) are different and are correlated due to the common term $h_r$ in their expressions, unlike a traditional BPSK system which has a single fading term.
\end{remark}

\subsection{Receiver Types}  \label{sec:RxTy}

The BER performance of the two receiver types $\mathcal{R}_1$ and $\mathcal{R}_2$, which correspond to the receivers with CSI and without CSI respectively, is analyzed in the paper. The first receiver $\mathcal{R}_1$ is assumed to track CSI perfectly which means the fading components $h_0$ and $h_1$ are known at the receiver. However, the complexity in the estimation of CSI may preclude some of these energy-constrained devices from tracking the channel, which is the primary motivation behind considering receiver $\mathcal{R}_2$ for which coding techniques such as differential coding are needed at the transmitter side to enable it to estimate data without CSI. In the absence of CSI, a receiver would not be able to map the conditional distributions of the received signal to the true message bit, thereby resulting in a poor decoding performance. We will elaborate on this point further in Section~\ref{sec:DetErrNoCSI}. With the help of differential encoding, receiver $\mathcal{R}_2$ will decode data bits by observing the change in two consecutive symbols rather from absolute values, thereby improving the BER performance of the receiver compared to an uncoded transmission.

Before going into further technical discussion, we define some key functional forms that will be used throughout this paper.  

\begin{definition} \label{def: x2rv}
The PDF of central chi-squared random variable $\chi^2(k)$ with degree $k$ is given by:
\begin{align}
f_{\chi^2}(x;k) &= \begin{cases}
\dfrac{x^{(\frac{k}{2}-1)} e^{-\frac{x}{2}}}{2^{\frac{k}{2}}\Gamma(\frac{k}{2})}, & x > 0,\\
0,        & \text{otherwise}.
\end{cases}
\end{align}
\end{definition}

\begin{definition} \label{def: rayrv}
The PDF of Rayleigh random variable with variance $\sigma^2$ of corresponding zero mean complex Gaussian RV is given by:
\begin{align}
f_{\rm Ray}(x;\sigma^2) &= \begin{cases}
\dfrac{2x}{\sigma^2} \exp\Big(-\dfrac{x^2}{\sigma^2}\Big), & x > 0,\\
0,        & \text{otherwise}.
\end{cases}
\end{align}
\end{definition}

\begin{definition} \label{def: FModBes1}
	
The modified Bessel function of the first kind with order $v$ is given by the expression:
\begin{align}
	I_v(z) &= (\frac{z}{2})^v \sum\limits_{i=0}^{\infty} \dfrac{(\frac{z^2}{4})^i}{i! \Gamma(v+i+1)},
\end{align}
and the corresponding integral form of the modified Bessel function when $v$ is an integer $n$ is given by:
\begin{align}
	I_n(z) &= \frac{1}{\pi} \int_{0}^{\pi} e^{z\cos\theta} \cos(n\theta) \mathrm{d}\theta.
\end{align}
\end{definition}

\begin{definition} \label{def: FModBes2}
	
The modified Bessel function of the second kind with order $v$ is given by the expression:
\begin{align}
	K_v(z) &= \frac{\pi}{2}  \dfrac{I_{-v}(z) - I_v(z)}{\sin v\pi},
\end{align}
where $I_v(z)$ is the modified Bessel function of the first kind.
\end{definition}

\begin{definition} \label{def: GMarcumQ}
	The generalized Marcum Q-function with degree $M$ and parameters $\alpha$, $\beta$ \cite{short12} is given by the expression:
	\begin{align}
		Q_M(\alpha,\beta) &= \dfrac{1}{\alpha^{M-1}} \int_{\beta}^{\infty} v^M \exp\Big(- \dfrac{v^2 + \alpha^2}{2} \Big) I_0(\alpha v)\, \mathrm{d}v.
	\end{align}	
\end{definition}

\section{Signal Detection} \label{sec:DetErrCSI}
In this section, we first study the detection process at receiver $\mathcal{R}_1$ in detail, beginning with the derivation of conditional distributions of the average signal energy represented by random variable $Y$ and the investigation of detection mechanisms to get the optimal detection threshold. We build on this analysis to study detection and error performance of receiver $\mathcal{R}_2$ focusing primarily on the elements differentiating the two receivers. 

\subsection{Receiver with CSI} \label{sec:RxCSI}

\subsubsection{Exact Distribution Functions} \label{sec:DistFunc}

The BTx node will modulate its own data onto the reflected ambient radio waves which means that the Rx node has to implement a mechanism to separate backscatter data from the \emph{ambient source} data. For this purpose, energy of the received signal is 
averaged over a window of $N$ samples. This mechanism results in a random variable (RV) $Y$ representing the average signal energy, and the operation is represented as follows \cite{shyam13}:
\begin{align}
Y = \frac{1}{N} \sum_{n=1}^{N} |y(n)|^{2} &= \frac{1}{N} \sum_{n=1}^{N} |(h_r +  \alpha h_{b} \text{ } b \text{ } h_t) x(n)  + w(n)|^{2} \label{eq:avgeeq}.
\end{align}

This problem is formulated as a binary hypothesis testing problem where the scenarios conditioned on bits $b=0$ and $b=1$ are taken as $\mathcal{H}_0$ (Null Hypothesis) as $\mathcal{H}_1$ (Alternate Hypothesis) respectively:
\begin{align}
	\mathcal{H}_0 : Y &= \frac{1}{N} \sum_{n=1}^{N} |h_0 x(n) + w(n)|^{2},& b = 0,\\
	\mathcal{H}_1 : Y &= \frac{1}{N} \sum_{n=1}^{N} |h_1 x(n) + w(n)|^{2},& b = 1.
\end{align}
The conditional probability density functions (PDFs) of $Y$ are crucial in the detection and estimation of the transmitted bit and are derived in the following Lemma.
\begin{lemma} \label{lem:NCX2_1}
The PDFs of $Y$ conditioned on $\mathcal{H}_0$, $\mu$ and $\mathcal{H}_1$, $\nu$ are respectively given by:
\begin{align}	
	f_{Y|\mathcal{H}_0,\mu}(t) &= \frac{2N}{\sigma^2} \sum \limits_{i=0}^{\infty} \dfrac{e^{-\frac{\mu N \bar{E}}{\sigma^2}} \left(\frac{\mu N \bar{E}}{\sigma^2}\right)^{i}}{i!} f_{\chi^2}(\frac{2N}{\sigma^2} t; 2N+2i) \label{eq:condpdf1},\\
	f_{Y|\mathcal{H}_1,\nu}(t) &= \frac{2N}{\sigma^2} \sum \limits_{i=0}^{\infty} \dfrac{e^{-\frac{\nu N \bar{E}}{\sigma^2}} \left(\frac{\nu N \bar{E}}{\sigma^2}\right)^{i}}{i!} f_{\chi^2}(\frac{2N}{\sigma^2} t; 2N+2i) \label{eq:condpdf2}.
\end{align}
\end{lemma}

\begin{IEEEproof}
	See Appendix~\ref{app:NCX2_1}.
\end{IEEEproof}

\begin{remark}\label{rem: condpdf}
	It can be observed that the PDFs of $Y$ conditioned on $\mathcal{H}_0$ and $\mathcal{H}_1$ are respectively dependent only on parameters $\mu$ and $\nu$, which are the squares of absolute values of the respective channel coefficients $h_0$ and $h_1$. Thus, the average BER can be written as the expectation of BER conditioned jointly (since they are not independent) on just $\mu$ and $\nu$.
\end{remark}

\subsubsection{Comparison with Approximate Distribution Functions} \label{sec:CompExactAppx}

The exact conditional PDFs derived here are compared with the approximations available in the literature. An alternate representation of $Y$ can be derived by expanding~\eqref{eq:avgeeq} and is given by the expression:
\begin{align}
	Y &= \frac{1}{N} \sum_{n=1}^{N} |y(n)|^{2} = \frac{1}{N} \sum_{n=1}^{N} y(n) y^*(n) \\
	&=  |h_r +  \alpha h_{b} h_t b|^2 \frac{1}{N} \sum_{n=1}^{N}  |x(n)|^2 + \frac{2}{N} \operatorname{Re} \left\{\left(h_r +  \alpha h_{b} h_t b\right) \sum_{n=1}^{N} x(n) w^*(n) \right\} + \frac{1}{N} \sum_{n=1}^{N}  |w(n)|^2\\
	&= \underbrace{|h_r +  \alpha h_{b} h_t b|^2 \bar{E}}_{\text{constant}}+\underbrace{\frac{2}{N} \operatorname{Re} \left\{\left(h_r +  \alpha h_{b} h_t b\right) \sum_{n=1}^{N} x(n) w^*(n) \right\}}_{\text{Gaussian RV}} + \underbrace{\frac{1}{N} \sum_{n=1}^{N}  |w(n)|^2}_{\text{Central-$\chi^2$ RV }}. \label{eq:sumexp}
\end{align}

\begin{figure}
    \centering
    \begin{subfigure}[b]{0.45\textwidth}
        \centering
        \includegraphics [width=\linewidth]{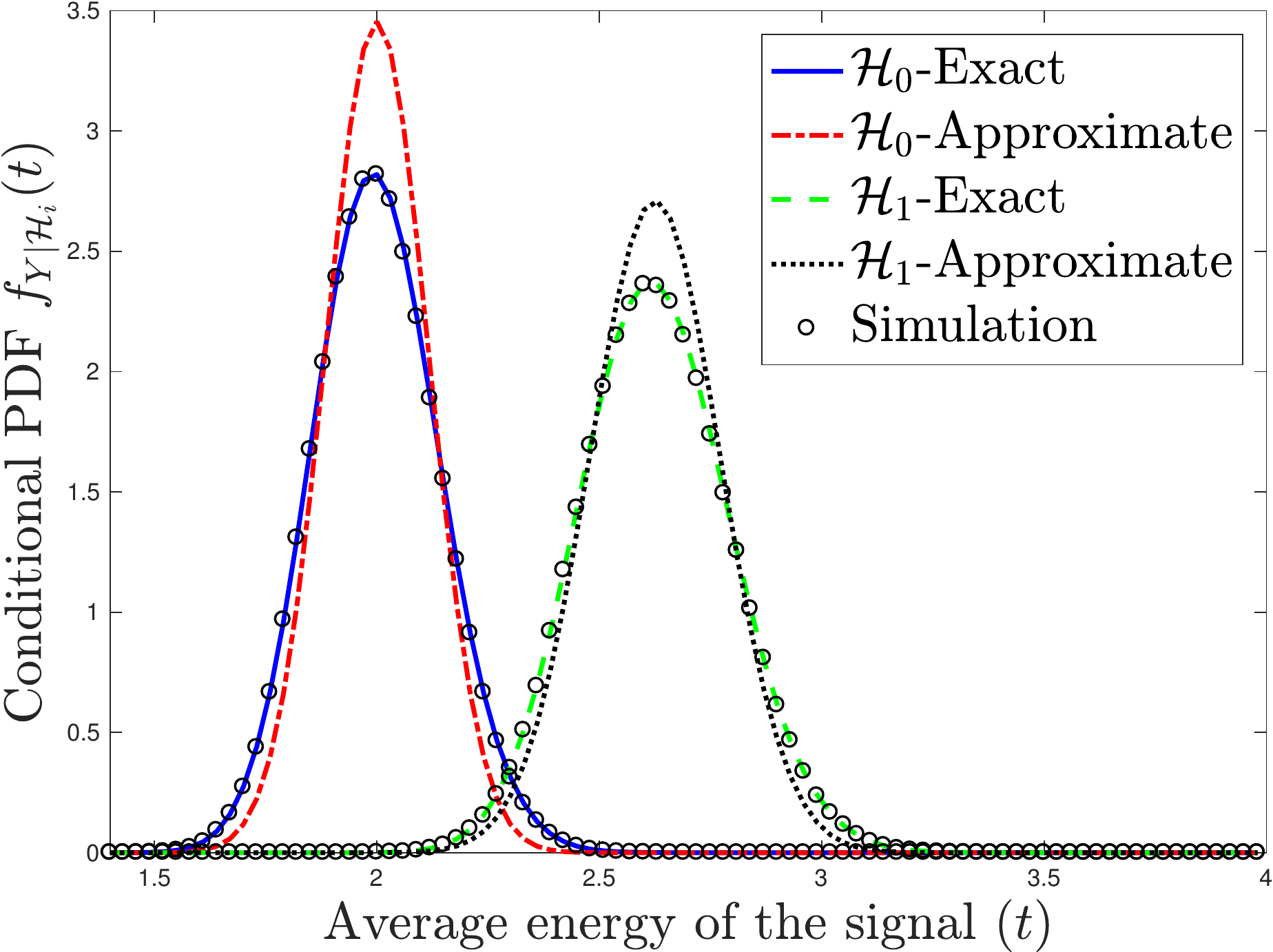}
         \caption{}\label{fig:ncx2_pdf_2a}
    \end{subfigure}
    ~ 
    \begin{subfigure}[b]{0.45\textwidth}
        \centering
        \includegraphics [width=\linewidth]{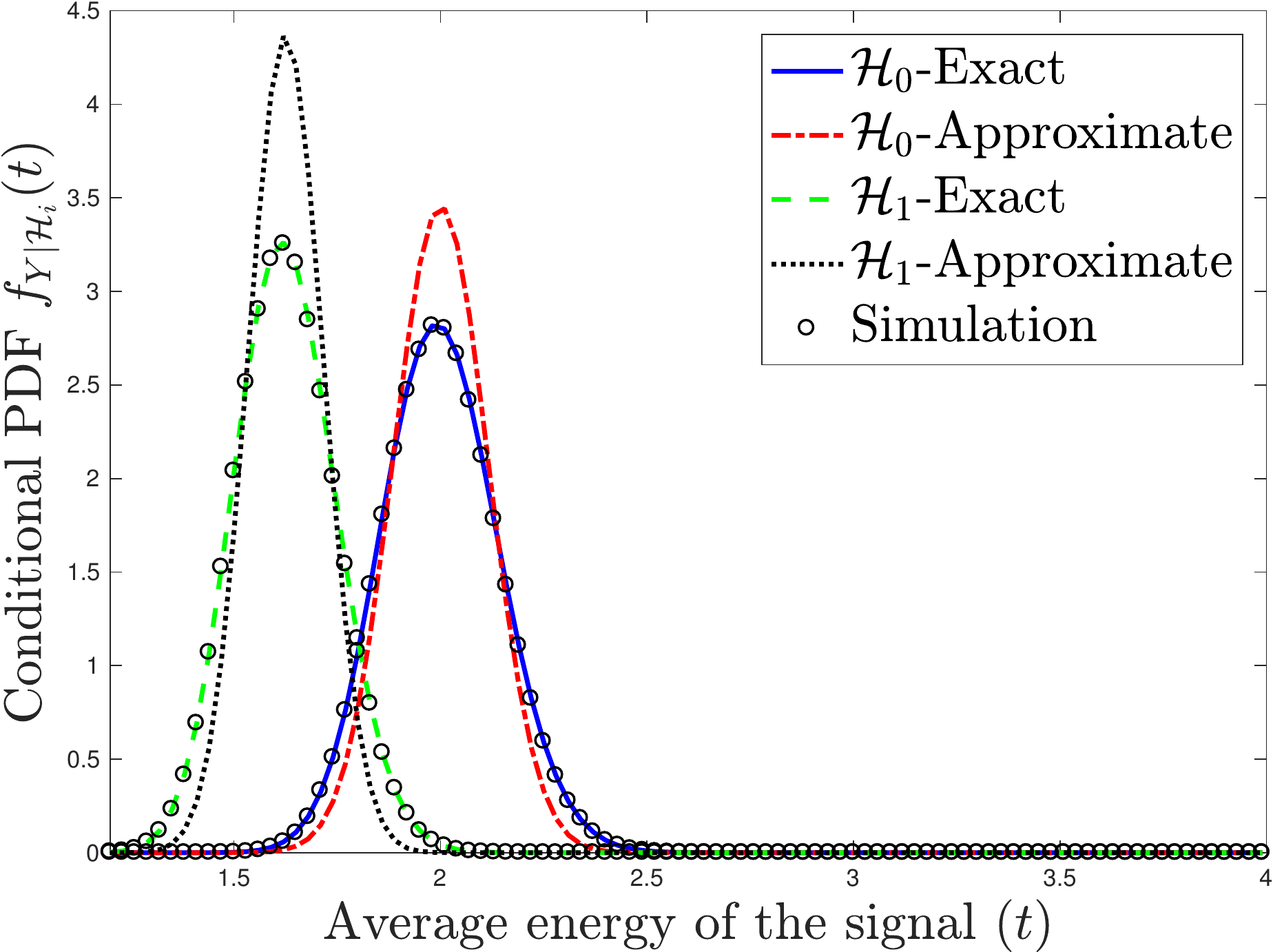}
	\caption{}\label{fig:ncx2_pdf_2b}
    \end{subfigure}
    \caption{Comparison of exact (derived in this paper) and approximate conditional PDFs \cite{chintha16} of average signal energy $Y$ for $\mu =1, \nu = 1.625$ (left) and $\mu =1, \nu = 0.625$ (right) at SNR = 0 dB, $N =150$.} \label{fig:ncx2_pdf_2}
\end{figure}

\begin{figure}
    \centering
    \begin{subfigure}[b]{0.45\textwidth}
        \centering
        \includegraphics [width=\linewidth]{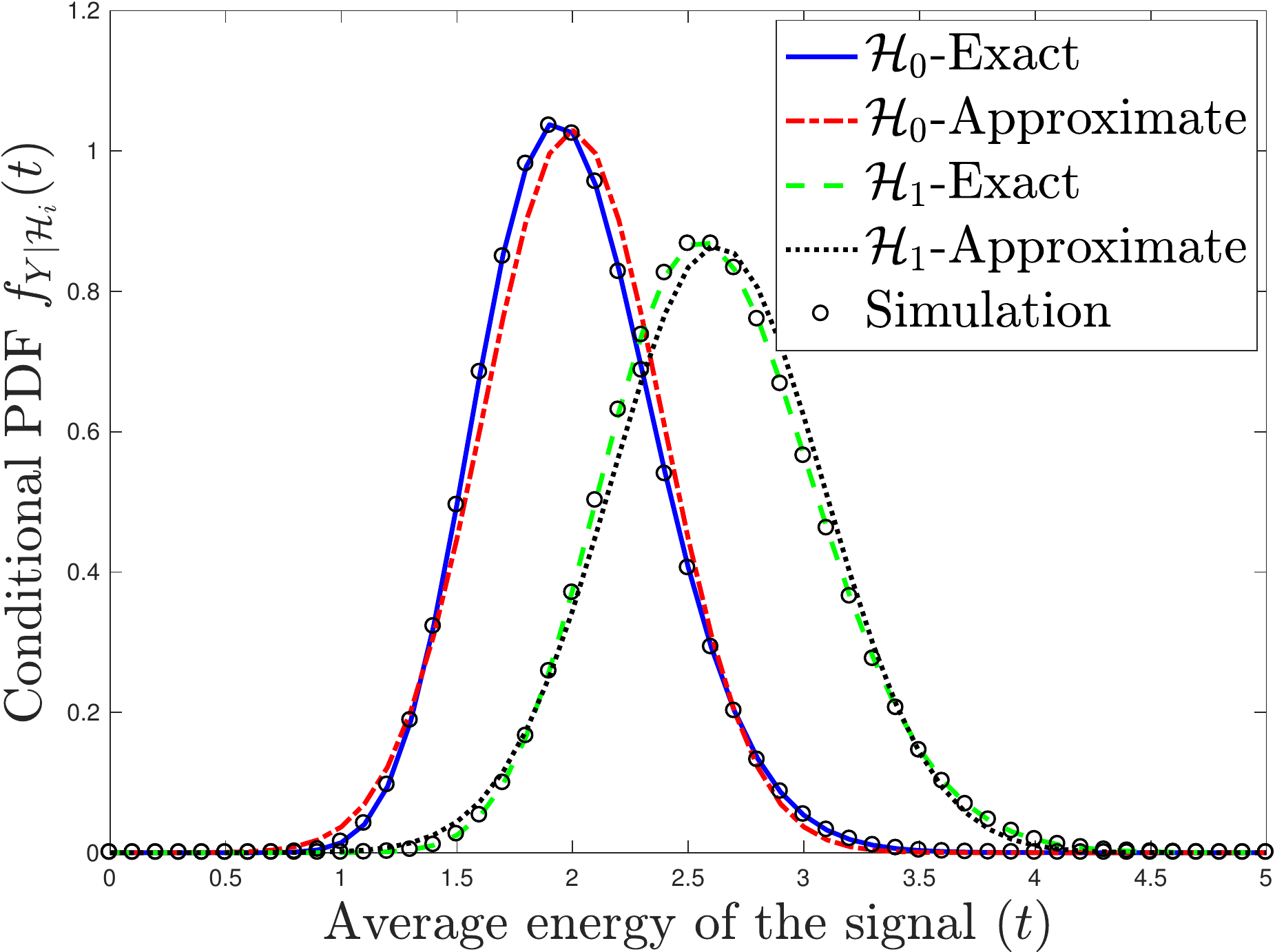}
         \caption{}\label{fig:ncx2_pdf_1a}
    \end{subfigure}
    ~ 
    \begin{subfigure}[b]{0.45\textwidth}
        \centering
        \includegraphics [width=\linewidth]{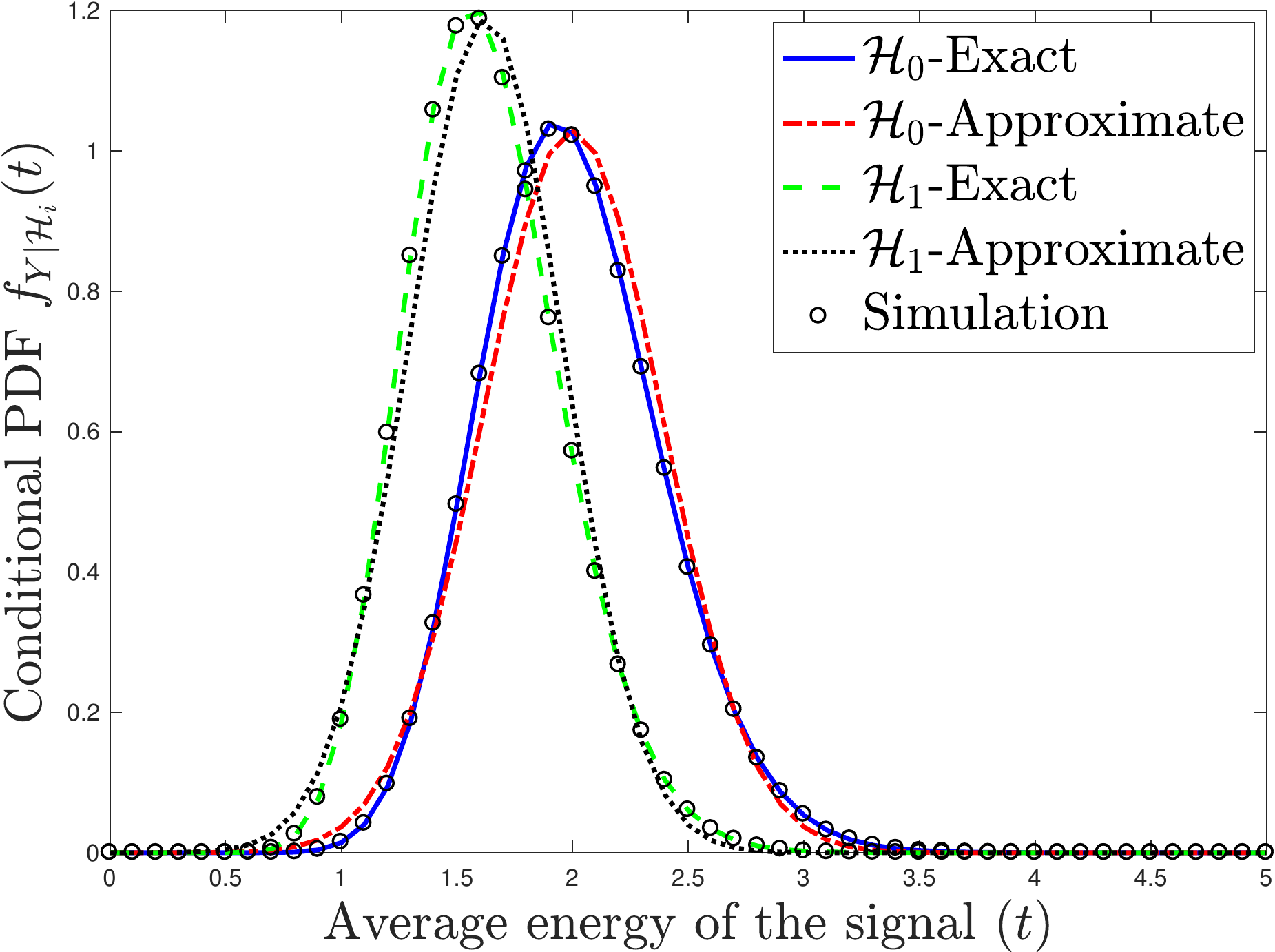}
	\caption{}\label{fig:ncx2_pdf_1b}
    \end{subfigure}
    \caption{Comparison of exact and approximate conditional PDFs \cite{chintha16} of average signal energy $Y$ for (a) $\mu =1, \nu = 1.625$ (b) $\mu =1, \nu = 0.625$ (SNR = 0 dB, $N =20$).}\label{fig:ncx2_pdf_1}
\end{figure}

One Gaussian approximation of $Y$ can be made by approximating the Central-$\chi^2$ RV with its mean value. This approximation is equivalent to the  approximation given for a large value of $N$ in \cite{chintha16}, which is also the preferred mode of approximation in the referenced paper. The exact and approximated conditional PDFs of the average signal energy $Y$ for $N$ = 150 and SNR = 0 dB are compared in Fig. \ref{fig:ncx2_pdf_2}, and the deviation in the plots is clearly noticeable. As expected, the exact distributions derived in this paper match exactly with the simulated conditional PDFs. On the other hand, the second Gaussian approximation of $Y$ can be done by approximating the Central-$\chi^2$ RV with a Gaussian RV of same mean and variance values. This approximation corresponds to the  approximation given for a small value of $N$ in \cite{chintha16}. The plots of the exact and approximated conditional PDFs in Fig. \ref{fig:ncx2_pdf_1} show that the approximations work reasonably well. The reason for the deteriorating performance of the approximations in \cite{chintha16} with increasing value of $N$ is due to the approximation of the aforementioned Central-$\chi^2$ RV with its mean value at higher values of $N$ which does not approximate the distribution properly. We observed that the approximation of this Central-$\chi^2$ RV instead with Gaussian (also proposed in \cite{chintha16} for smaller $N$) works better for all values of $N$. From this point onward, the two approximations are referred to as the first and second Gaussian approximation of the conditional PDFs of $Y$.

The impact of channel variations on the conditional PDFs is analyzed by plotting them for the two sets of values of channel parameters $\mu$ and $\nu$. When the two sub-plots in Figs. \ref{fig:ncx2_pdf_2} and \ref{fig:ncx2_pdf_1} are compared, the conditional distributions of two hypotheses are observed to interchange their positions which means that the relative positions of the conditional distributions of two hypotheses change with channel parameters $\mu$ and $\nu$. Further, as the value of sample length $N$ increases the conditional variance of $Y$ decreases and this results in the concentration of the conditional PDFs. This effect can be observed in Figs. \ref{fig:ncx2_pdf_2} and \ref{fig:ncx2_pdf_1} by checking the difference in the supports over which the PDFs are mainly concentrated.

\subsubsection{Detection Threshold} \label{sec:DetTh}

In the optimal detection of the standard BPSK modulation using MLT detection, the threshold value calculated as the intersection point of conditional likelihood functions has a tractable solution. However, the solution for the MLT estimation in an ambient backscatter system is intractable, and for this reason, we consider two other strategies called MT detection and approximate MLT detection along with the optimal MLT detection.
\paragraph{Mean Threshold (MT) Detection}  \label{sec:MTDet}
The threshold value of MT detection method is evaluated as the mean of the conditional expectations of average signal energy $Y$ given $\mathcal{H}_0,\mu$ and $\mathcal{H}_1,\nu$:
\begin{align}
T_{\rm mt} &= \frac {\mathbb{E}[Y|\mathcal{H}_0] + \mathbb{E}[Y|\mathcal{H}_1]} {2} = \sigma^2 + \frac{\bar{E}( \mu + \nu)}{2}.
\end{align}

\paragraph{Maximum Likelihood Threshold (MLT) Detection}  \label{sec:OptDet}

Here, we derive the expression of optimal threshold value for maximum likelihood detection. The representation of the conditional PDFs of $Y$ in terms of ``sums of terms'' as given in~\eqref{eq:condpdf1} and~\eqref{eq:condpdf2} can be modified to an alternate integral form using the modified Bessel function of first kind which is given for any integer order. Using this integral representation we can derive expression for the MLT, which unfortunately however does not have a tractable form. This result is presented in the following Lemma. 

\begin{lemma} \label{lem:MLthold}
	The optimal detection rule or threshold $T_{\rm mlt}$ (a function of $\mu$ and $\nu$) is calculated by solving the expression given below:
\begin{align}	
	&\frac{N}{\sigma^2}  e^{-\left(\frac{N}{\sigma^2} T_{\rm mlt}+\frac{N \mu \bar{E}}{\sigma^2}\right)} \left(\dfrac{4T_{\rm mlt}}{\mu  \bar{E}}\right)^{\frac{N-1}{2}} I_{N-1}\left(\dfrac{2N}{\sigma^2} \sqrt{\mu  \bar{E} T_{\rm mlt}}\right) \nonumber\\
	&= \frac{N}{\sigma^2}  e^{-\left(\frac{N}{\sigma^2}T_{\rm mlt}+\frac{N \nu \bar{E}}{\sigma^2}\right)} \left(\dfrac{4T_{\rm mlt}}{\nu  \bar{E}}\right)^{\frac{N-1}{2}} I_{N-1}\left(\dfrac{2N}{\sigma^2} \sqrt{\nu  \bar{E} T_{\rm mlt}}\right), \label{eq:mleq}
\end{align}
and the expression can be simplified as follows:
\begin{align}
	e^{\frac{N}{\sigma^2}\bar{E}(\nu-\mu)} \left(\frac{\nu}{\mu}\right)^{\frac{N-1}{2}} \int_{0}^{\pi} e^{\frac{2N}{\sigma^2} \sqrt{\mu \bar{E} \, T_{\rm mlt}} \cos\theta} \cos (N-1)\theta\, \mathrm{d}\theta &= \int_{0}^{\pi} e^{\frac{2N}{\sigma^2} \sqrt{\nu \bar{E} \, T_{\rm mlt}} \cos\theta}  \cos (N-1)\theta\, \mathrm{d}\theta. \label{eq:mleq2}
\end{align}
\end{lemma}

\begin{IEEEproof}
	See Appendix~\ref{app:MLthold}.
\end{IEEEproof}

\paragraph{ Approximate MLT Detection}  \label{sec:OptDet}
As discussed above, solving~\eqref{eq:mleq2} gives the optimal ML threshold value. The presence of $T_{\rm mlt}$, the variable we are evaluating, inside the integral makes the problem highly intractable and the procedure is not so straightforward. To simplify the computations, we provide approximate solutions using Gaussian approximations of the conditional PDFs that we discussed earlier. We remind again that the selection of threshold value is an independent process from the characterization of signal distributions. The conditional distributions derived in subsection \ref{sec:DistFunc} are exact without any approximations as mentioned in the contributions of this paper. The approximations of the conditional distributions is only used for the derivation of tractable solutions to MLT to enable faster numerical computations. These approximate MLT thresholds are similar to the ones used in \cite{chintha16}. 

\begin{lemma} \label{lem:APXMLthold}
	The approximate ML thresholds $T_{\rm mlt, app 1}$ and $T_{\rm mlt, app 2}$ for the two Gaussian approximations of conditional distributions of $Y$ are given by the following expressions:
\begin{align}
	T_{\rm mlt, app 1} &= \sigma^2 + \sqrt{\mu\nu \bar{E}\Big(\frac{2\sigma^2}{N(\nu - \mu)} \ln\big(\frac{\nu}{\mu}\big) + \bar{E}\Big)},\\
	T_{\rm mlt, app 2} &=  \frac{\sigma^2}{2} + \sqrt{\frac{\sigma^4}{4} + \mu \nu \bar{E}^2 + \frac{ \mu + \nu}{2} \bar{E} \sigma^2 + \frac{\left(2\mu\bar{E}+\sigma^2\right) \left(2\nu\bar{E}+\sigma^2\right)\sigma^2}{2 N (\nu-\mu) \bar{E}} \ln \left( \frac{2\nu\bar{E}+\sigma^2}{2\mu\bar{E}+\sigma^2}\right)}.
\end{align}
\end{lemma}

\begin{IEEEproof}
	See Appendix~\ref{app:APXMLthold}.
\end{IEEEproof}

\subsection{Receiver without CSI} \label{sec:DetErrNoCSI}

The assumption of CSI tracking at receiver $\mathcal{R}_1$ gave one the freedom to choose different evaluation strategies in estimating the threshold value. For energy-constrained devices like sensors, tracking a channel continuously may not be the ideal use of their energy and would be beneficial if detection mechanisms without (or partial) channel information can be implemented. By partial channel information, we mean that there is some measure of the channel like mean energy of the channel estimates. Additionally, energy constraints in some of the devices restrict the evaluation of complex numerical operations inhibiting the implementation of most of the threshold techniques. These are the primary factors motivating the pursuit of detection schemes in a receiver without (or partial) channel estimates which can result in reasonable performance. 

The fading components in the binary hypothesis problem formulated earlier in~\eqref{eq: BinHyp} are observed to be different under each hypothesis. Both of the fading terms are complex and the magnitude of one component can be either smaller or bigger than the other component. As observed in the analysis related to conditional distributions, the conditional PDFs interchange positions with respect to the relative values of these components. Without information on the relative location of the conditional distributions of the two hypotheses, the threshold detector can incorrectly map the received average signal to a different hypothesis with high probability. To overcome the ambiguity of mapping correct conditional PDFs at receiver $\mathcal{R}_2$, differential encoding is implemented at the transmitter which reduces the complexity of the receiver \emph{albeit} with a slight degradation in error performance \cite{shyam13}. Mathematically, the output of a differential encoding block is given by:
\begin{align}
b(n) = b(n-1) \oplus m(n), 
\end{align}
where $\oplus$ is the exclusive-or (XOR) operation, $b(n)$ is the transmitted bit at current time instant, $b(n-1)$ is the bit transmitted in previous time instant, and $m(n)$ is the message bit to be transmitted in the current time instant. At the receiver, $m(n)$ can be decoded with a similar XOR operation given by: 
\begin{align}
\hat{m}(n) = \hat{b}(n) \oplus \hat{b}(n-1),
\end{align}
where $\hat{b}(n)$ and $\hat{b}(n-1)$ are the symbols received at the current and previous time instants respectively. It can be observed that the information in differential encoding is encoded as a change rather than absolute values of the transmitted symbols, and in the differential decoding block at the receiver two consecutive symbols are used to detect each bit in the stream. Since the differential decoding takes in two consecutive symbols at a time, the value of fading coefficient is assumed to be the same over the two symbols (fairly reasonable assumption).

\paragraph*{Threshold Strategies}
As there is no channel information at $\mathcal{R}_2$, we can only use threshold techniques which do not involve the explicit estimation of the channel state. This is where the simplicity of evaluating the Mean threshold (MT) allows one to employ the technique for this receiver. The threshold of MT detection can be implemented in practice by averaging the energy of samples received over the first few time slots in the channel coherence period. 

\section{Bit Error Rate Analysis}  \label{sec:BERRx}

In this section, we analyze the performance of the detection strategies by evaluating the BER expressions. The conditional BER expressions are also evaluated in terms of the generalized Marcum Q-function similar to the accepted representation of BER of the Gaussian distributed signals using the standard Q-function. This form of presentation of the conditional BER allows us to show the dependence of optimal BER performance on the SNR of the ambient signal which is demonstrated in the next subsection.

As noted in Remark \ref{rem: condpdf}, the average BER of an ambient backscatter system is dependent on joint distribution of the fading components $\mu$ and $\nu$. The analytical expression of the average BER in a fading channel can be written as:
\begin{align}
P_{e} &= \mathbb{E}_{\mu, \nu} [P(e|\mu, \nu)]\\
&= \int_{0}^{\infty} \int_{0}^{\infty} f_{\mu, \nu}(\mu, \nu) P(e|\mu, \nu) \, \mathrm{d}\nu \, \mathrm{d}\mu \label{Eq: berexp},
\end{align}	
where $f_{\mu, \nu}(\mu, \nu)$ is the joint probability density of fading components $\mu$ and $\nu$, and $P(e|\mu, \nu)$ is the error probability conditioned on $\mu$ and $\nu$. To the best of our understanding, existing works do not deal with the characterization of this joint probability density and hence the average BER analysis for this setup.

\subsection{Conditional Error Probability}  \label{sec:condBER}

First, we derive the expressions of conditional error probabilities for receiver $\mathcal{R}_1$ and then extend the analysis to receiver $\mathcal{R}_2$. The conditional error probability $P(e|\mu, \nu)$ of a receiver is given by the expression:
\begin{align}
P(e|\mu, \nu) &= P(\mathcal{H}_0) P(e| \mathcal{H}_0, \mu) + P(\mathcal{H}_1) P(e| \mathcal{H}_1, \nu).
\end{align}
Assuming the symbols are equally likely, the prior probabilities of the two hypotheses are given by $P(\mathcal{H}_0) = P(\mathcal{H}_1) = \frac{1}{2}$. The conditional error probability of each hypothesis of receiver $\mathcal{R}_1$ is given by the following relation since the relative values of $\mu$ and $\nu$ change the relative positions of the conditional distribution curves:
\begin{align}
 P_{\mathcal{R}_1}(e| \mathcal{H}_0, \mu)  &= \begin{cases}
  \int\limits_{0}^{T(\mu, \nu)} f_{Y|\mathcal{H}_0,\mu}(t) \, \mathrm{d}t, & \nu < \mu, \\
  \int\limits_{T(\mu, \nu)}^{\infty}f_{Y|\mathcal{H}_0,\mu}(t) \, \mathrm{d}t, & \nu \ge \mu.
 \end{cases}\\
  P_{\mathcal{R}_1}(e| \mathcal{H}_1, \nu)  &= \begin{cases}
  \int\limits_{T(\mu, \nu)}^{\infty}f_{Y|\mathcal{H}_1,\nu}(t) \, \mathrm{d}t, & \nu < \mu, \\
   \int\limits_{0}^{T(\mu, \nu)}f_{Y|\mathcal{H}_1,\nu}(t) \, \mathrm{d}t, & \nu \ge \mu.
 \end{cases}
 \end{align}
 
When $\nu \ge \mu$, analytical expression of the conditional bit error rate is given by:
\begin{align}
P^1_{\mathcal{R}_1}(e|\mu, \nu) &=  P(\mathcal{H}_0) P_{\mathcal{R}_1}(e| \mathcal{H}_0, \mu) + P(\mathcal{H}_1) P_{\mathcal{R}_1}(e| \mathcal{H}_1, \nu)\\
&= \frac{1}{2} \left(\int_{T}^{\infty} f_{Y|\mathcal{H}_0,\mu}(t) \, \mathrm{d}t + \int_{0}^{T} f_{Y|\mathcal{H}_1,\nu}(t) \, \mathrm{d}t\right).\label{eq:erprob}
\end{align}

On the other hand for $\nu < \mu$, the conditional bit error rate is given by:
\begin{align}
P^2_{\mathcal{R}_1}(e|\mu, \nu) &=  P(\mathcal{H}_0) P_{\mathcal{R}_1}(e| \mathcal{H}_0, \mu) + P(\mathcal{H}_1) P_{\mathcal{R}_1}(e| \mathcal{H}_1, \nu)\\
&= \frac{1}{2} \left(\int_{0}^{T} f_{Y|\mathcal{H}_0,\mu}(t) \, \mathrm{d}t + \int_{T}^{\infty} f_{Y|\mathcal{H}_1,\nu}(t) \, \mathrm{d}t\right) = 1 - P^1_{\mathcal{R}_1}(e|\mu, \nu).
\end{align}

The value of conditional BER is a function of  the instantaneous values of parameters $\mu$ and $\nu$ and can take either $P_{\mathcal{R}_1}(e|\mu, \nu)  = P^1_{\mathcal{R}_1}(e|\mu, \nu)$ or $P_{\mathcal{R}_1}(e|\mu, \nu) (e|\mu, \nu) = P^2_{\mathcal{R}_1}(e|\mu, \nu)$ depending on the relative values of the two parameters. When differential encoding is implemented at transmitter for receiver $\mathcal{R}_2$, the conditional BER expression simplifies to a single expression. For the receiver $\mathcal{R}_2$, error is going to occur at the output of differential decoding when only one of the two consecutive bits of the received symbols flips. Also, observe that both of the detected bits are independent which simplifies the analysis, and we can write the expression of the conditional BER as follows:
\begin{align}
P_{\mathcal{R}_2}(e|\mu, \nu) &= P(\hat{Y}_k \neq Y_k) P(\hat{Y}_{k-1} = Y_{k-1}) + P(\hat{Y}_k = Y_k) P(\hat{Y}_{k-1} \neq Y_{k-1})\\
&= 2 P(\hat{Y}_k \neq Y_k) P(\hat{Y}_{k-1} = Y_{k-1})\\
&= 2 P^1_{\mathcal{R}_1}(e|\mu, \nu) P^2_{\mathcal{R}_1}(e|\mu, \nu).
\end{align}

The Marcum Q-function is extensively used as a cumulative distribution function for noncentral chi, noncentral chi-squared and Rice distributions and many algorithms for efficient evaluation of the function are implemented in hardware and software. Hence, it would be highly beneficial to give equivalent representations of the conditional BER in terms of Marcum Q-function.

The conditional error probabilities of the two receivers $\mathcal{R}_1$ and $\mathcal{R}_2$ of ambient backscatter systems in terms of the generalized Marcum Q-function can be expressed as \cite{short12,shnidman89} :
\begin{align}
&P_{\mathcal{R}_1}(e|\mu, \nu) = \begin{cases}
  P^1_{\mathcal{R}_1}(e|\mu, \nu)   & \nu < \mu, \\
  P^2_{\mathcal{R}_1}(e|\mu, \nu)   & \nu \ge \mu.
 \end{cases}\\
 &= \begin{cases}
  \frac{1}{2} \left\{1 + Q_N\left(\sqrt{2N \frac{\mu\bar{E} }{\sigma^2}}, \sqrt{ 2N \frac{T(\mu,\nu)}{\sigma^2}}\right) - Q_N\left(\sqrt{2N \frac{\nu \bar{E} }{\sigma^2}}, \sqrt{2N \frac{ T(\mu,\nu)}{\sigma^2}}\right)\right\}  & \nu < \mu, \\
\frac{1}{2} \left\{1+ Q_N\left(\sqrt{2N \frac{ \nu\bar{E} }{\sigma^2}}, \sqrt{2N \frac{ T(\mu,\nu)}{\sigma^2}}\right) -Q_N\left(\sqrt{2N  \frac{\mu\bar{E} }{\sigma^2}}, \sqrt{2N \frac{T(\mu,\nu)}{\sigma^2}}\right) \right\} & \nu \ge \mu.
 \end{cases}\label{eq:marcumqnc}
 \end{align}
 \begin{align}
&P_{\mathcal{R}_2}(e|\mu, \nu) =  2 P^1_{\mathcal{R}_1}(e|\mu, \nu) P^2_{\mathcal{R}_1}(e|\mu, \nu)\\
	&= \frac{1}{2} \left\{1 + Q_N\left(\sqrt{2N \frac{\mu\bar{E} }{\sigma^2}}, \sqrt{ 2N \frac{T(\mu,\nu)}{\sigma^2}}\right) - Q_N\left(\sqrt{2N  \frac{\nu\bar{E} }{\sigma^2}}, \sqrt{2N \frac{ T(\mu,\nu)}{\sigma^2}}\right)\right\} \times\nonumber\\ 
	& \left\{1+ Q_N\left(\sqrt{2N  \frac{\nu\bar{E} }{\sigma^2}}, \sqrt{2N \frac{ T(\mu,\nu)}{\sigma^2}}\right) -Q_N\left(\sqrt{2N \frac{\mu \bar{E} }{\sigma^2}}, \sqrt{2N \frac{T(\mu,\nu)}{\sigma^2}}\right) \right\} \\
	&= \frac{1}{2}  - \frac{1}{2} \left\{Q_N\left(\sqrt{2N \frac{\nu\bar{E} }{\sigma^2}}, \sqrt{2N \frac{ T(\mu,\nu)}{\sigma^2}}\right) - Q_N\left(\sqrt{2N  \frac{ \mu\bar{E} }{\sigma^2}}, \sqrt{2N\frac{T(\mu,\nu)}{\sigma^2}}\right)\right\}^2. \label{eq:marcumqdc}
\end{align}

\begin{remark}\label{rem:BERexp}
We can observe from~\eqref{eq:marcumqnc} and~\eqref{eq:marcumqdc} that the conditional BER expressions are functions of the parameters $N, \dfrac{\mu\bar{E}} {\sigma^2}, \dfrac{\nu\bar{E}} {\sigma^2}$ and $\dfrac{T(\mu,\nu)} {\sigma^2}$. The fractions $\dfrac{T(\mu,\nu)} {\sigma^2}$ for the MT threshold and the two approximate MLTs threshold techniques can be modified as: 
\begin{align}
\frac{T_{\rm mt}}{\sigma^2} &=  1 + \frac{\mu + \nu}{2} \frac{\bar{E}}{\sigma^2} = 1 + \frac{ \frac{ \mu\bar{E}}{\sigma^2}+  \frac{\nu\bar{E}}{\sigma^2}}{2} , \\
\frac{T_{\rm mlt, app 1}}{\sigma^2} &= 1 + \sqrt{\mu\nu \frac{\bar{E}}{\sigma^2}\left(\frac{2}{N(\nu - \mu)} \ln\left(\frac{\nu}{\mu}\right) + \frac{\bar{E}}{\sigma^2}\right)}\\
&= 1 + \sqrt{ \frac{\nu\bar{E}}{\sigma^2}\left(\frac{2\frac{\mu \bar{E}}{\sigma^2}}{N\left(\frac{\nu \bar{E}}{\sigma^2} - \frac{\mu \bar{E}}{\sigma^2}\right)} \ln\left(\frac{\frac{\nu \bar{E}}{\sigma^2}}{\frac{\mu \bar{E}}{\sigma^2}}\right) + \frac{\mu \bar{E}}{\sigma^2}\right)},\\
\frac{T_{\rm mlt, app 2}}{\sigma^2} &= \frac{1}{2} + \sqrt{\frac{1}{4} + \mu \nu \left(\frac{\bar{E}}{\sigma^2}\right)^2 + \frac{ \mu + \nu}{2} \frac{\bar{E}}{\sigma^2} + \frac{\left(2\mu \frac{\bar{E}}{\sigma^2} +1\right) \left(2\nu\frac{\bar{E}}{\sigma^2}+ 1\right)}{2 N (\nu-\mu) \frac{\bar{E}}{\sigma^2}} \ln \left( \frac{2\nu\frac{\bar{E}}{\sigma^2}+1}{2\mu\frac{\bar{E}}{\sigma^2}+1}\right)}\\
&= \frac{1}{2} + \sqrt{\frac{1}{4} + \frac{\nu \bar{E}}{\sigma^2} \frac{\mu \bar{E}}{\sigma^2} + \frac{ \frac{ \mu\bar{E}}{\sigma^2}+  \frac{\nu\bar{E}}{\sigma^2}}{2} + \frac{\left(\frac{2\mu\bar{E}}{\sigma^2} +1\right) \left(\frac{2\nu\bar{E}}{\sigma^2}+ 1\right)}{2 N \left(\frac{\nu\bar{E}}{\sigma^2}-\frac{\mu\bar{E}}{\sigma^2}\right) } \ln \left( \frac{\frac{2\nu\bar{E}}{\sigma^2}+1}{\frac{2\mu\bar{E}}{\sigma^2}+1}\right)},
\end{align}
which are functions of the other three parameters $N, \dfrac{\mu\bar{E}} {\sigma^2}$ and $\dfrac{\nu\bar{E}} {\sigma^2}$.

Even though MLT technique does not have a closed form expression for the threshold, we show that the solution $\dfrac{T_{\rm mlt}} {\sigma^2}$ of~\eqref{eq:mleq2} has to be a function of the same three parameters. The rearranged form of~\eqref{eq:mleq2} given below is a function of the three parameters $N, \dfrac{\mu\bar{E}} {\sigma^2}$ and $\dfrac{\nu\bar{E}} {\sigma^2}$ and hence, the solution $\dfrac{T_{\rm mlt}} {\sigma^2}$ of the equation would also be a function of the three parameters.
\begin{align}
e^{N\left(\frac{\nu\bar{E}}{\sigma^2}-\frac{\mu\bar{E}}{\sigma^2}\right)} \left|\frac{\frac{\nu\bar{E}}{\sigma^2}}{\frac{\mu\bar{E}}{\sigma^2}}\right|^{\frac{N-1}{2}} \int_{0}^{\pi} e^{2N \sqrt{ \frac{\mu\bar{E}}{\sigma^2} \, \frac{T_{\rm mlt}}{\sigma^2}} \cos\theta} \cos (N-1)\theta\, \mathrm{d}\theta &= \int_{0}^{\pi} e^{2N \sqrt{ \frac{\nu\bar{E}}{\sigma^2} \, \frac{T_{\rm mlt}}{\sigma^2}} \cos\theta}  \cos (N-1)\theta\, \mathrm{d}\theta.
\end{align}
The fractions $ \dfrac{\mu\bar{E}} {\sigma^2}$ and $\dfrac{\nu\bar{E}} {\sigma^2}$ are the received SNRs under the two hypotheses. Hence, it can be concluded that the conditional BER of the MT, approximate MLTs and the optimal MLT threshold mechanisms depend upon the signal and noise strengths through SNR and not their respective energies separately.
\end{remark}

\subsection{Average Error Probability}  \label{sec:avgBER}

The second component required in the average BER expression is the joint distribution function of fading components $\mu$ and $\nu$, which is derived in the following Lemma.

\begin{lemma} \label{lem:JointDist}
The joint density of the fading components $\mu$ and $\nu$ is given by the following expression:
\begin{align}
f_{\mu, \nu}(\mu, \nu)
&= \dfrac{1}{\pi \sigma_h^2} e^{-\frac{\mu}{\sigma_h^2}} \dfrac{1}{2\pi (|\alpha|\sigma_h^2)^2} \int_{0}^{2\pi} \int_{0}^{2\pi} K_0\Big(\dfrac{\sqrt{\mu +  \nu - 2 \sqrt{\mu \nu} \cos (\theta_{h_1} - \theta_{h_0})}}{\frac{|\alpha|\sigma_h^2}{2}}\Big) \mathrm{d}\theta_{h_1} \mathrm{d}\theta_{h_0} \label{Eq: jointdist},
\end{align}
where $K_0(z)$ is the zeroth order modified Bessel function of second kind.
\end{lemma}
\begin{IEEEproof}
	See Appendix~\ref{app:JointDist}.
\end{IEEEproof}

We can now provide the final result of the paper which quantifies the error performance of ambient backscatter systems in terms of the average BER. The following theorem gives the final average BER expressions for both receivers $\mathcal{R}_1$ and $\mathcal{R}_2$ in the ambient backscatter systems.
\begin{theorem} \label{thm:BERExp}
The average BER of the receivers $\mathcal{R}_1$ (with CSI) and $\mathcal{R}_2$ (without CSI) in an ambient backscatter system are respectively given by the expressions:
\begin{align}
&P_{\mathcal{R}_1}(e)\nonumber \\
&= \int_{\mu = 0}^{\infty} \int_{\nu =0}^{\mu} \dfrac{1}{\pi \sigma_h^2} e^{-\frac{\mu}{\sigma_h^2}} \dfrac{1}{2\pi (|\alpha|\sigma_h^2)^2} \int_{0}^{2\pi} \int_{0}^{2\pi} K_0\left(\dfrac{\sqrt{\mu +  \nu - 2 \sqrt{\mu \nu} \cos (\theta_{h_1} - \theta_{h_0})}}{\frac{|\alpha|\sigma_h^2}{2}}\right) \mathrm{d}\theta_{h_1} \mathrm{d}\theta_{h_0} \nonumber\\
&\times  \frac{1}{2} \left\{1 + Q_N\left(\sqrt{2N \frac{\mu\bar{E} }{\sigma^2}}, \sqrt{ 2N \frac{T(\mu,\nu)}{\sigma^2}}\right) - Q_N\left(\sqrt{2N \frac{\nu\bar{E} }{\sigma^2}}, \sqrt{2N \frac{ T(\mu,\nu)}{\sigma^2}}\right)\right\}  \, \mathrm{d}\nu \, \mathrm{d}\mu \nonumber \\
&+ \int_{\mu = 0}^{\infty} \int_{\nu =\mu}^{\infty} \dfrac{1}{\pi \sigma_h^2} e^{-\frac{\mu}{\sigma_h^2}} \dfrac{1}{2\pi (|\alpha|\sigma_h^2)^2} \int_{0}^{2\pi} \int_{0}^{2\pi} K_0\left(\dfrac{\sqrt{\mu +  \nu - 2 \sqrt{\mu \nu} \cos (\theta_{h_1} - \theta_{h_0})}}{\frac{|\alpha|\sigma_h^2}{2}}\right) \mathrm{d}\theta_{h_1} \mathrm{d}\theta_{h_0} \nonumber \\
&\times \frac{1}{2} \left\{1+ Q_N\left(\sqrt{2N \frac{ \nu\bar{E} }{\sigma^2}}, \sqrt{2N \frac{ T(\mu,\nu)}{\sigma^2}}\right) -Q_N\left(\sqrt{2N \frac{ \mu\bar{E} }{\sigma^2}}, \sqrt{2N \frac{T(\mu,\nu)}{\sigma^2}}\right) \right\}  \, \mathrm{d}\nu \, \mathrm{d}\mu,\\
&P_{\mathcal{R}_2}(e) \nonumber \\
&= \int_{\mu = 0}^{\infty} \int_{\nu =0}^{\infty} \dfrac{1}{\pi \sigma_h^2} e^{-\frac{\mu}{\sigma_h^2}} \dfrac{1}{2\pi (|\alpha|\sigma_h^2)^2} \int_{0}^{2\pi} \int_{0}^{2\pi} K_0\left(\dfrac{\sqrt{\mu +  \nu - 2 \sqrt{\mu \nu} \cos (\theta_{h_1} - \theta_{h_0})}}{\frac{|\alpha|\sigma_h^2}{2}}\right) \mathrm{d}\theta_{h_1} \mathrm{d}\theta_{h_0} \nonumber\\
&\times  \left(\frac{1}{2}  - \frac{1}{2} \left\{Q_N\left(\sqrt{2N \frac{\nu\bar{E} }{\sigma^2}}, \sqrt{2N \frac{ T(\mu,\nu)}{\sigma^2}}\right) - Q_N\left(\sqrt{2N \frac{ \mu \bar{E} }{\sigma^2}}, \sqrt{2N\frac{T(\mu,\nu)}{\sigma^2}}\right)\right\}^2\right)  \, \mathrm{d}\nu \, \mathrm{d}\mu,
\end{align}
where $T(\mu,\nu)$ is the threshold value which depends on the employed detection strategy.
\end{theorem}

\begin{IEEEproof}
Using the definition of average BER in~\eqref{Eq: berexp} of an ambient backscatter system, the equivalent expression for receiver $\mathcal{R}_1$ is given by:
\begin{align}
P_{\mathcal{R}_1}(e) &= \int_{0}^{\infty} \int_{0}^{\infty} f_{\mu, \nu}(\mu, \nu) P_{\mathcal{R}_1}(e|\mu, \nu) \, \mathrm{d}\nu \, \mathrm{d}\mu \\
&\stackrel{(k)}{=}  \int_{0}^{\infty} \int_{0}^{\mu} f_{\mu, \nu}(\mu, \nu) P^1_{\mathcal{R}_1}(e|\mu, \nu) \, \mathrm{d}\nu \, \mathrm{d}\mu + \int_{0}^{\infty} \int_{\mu}^{\infty} f_{\mu, \nu}(\mu, \nu) P^2_{\mathcal{R}_1}(e|\mu, \nu) \, \mathrm{d}\nu \, \mathrm{d}\mu,
\end{align}
where $(k)$ follows from the piecewise expressions of $P_{\mathcal{R}_1}(e|\mu, \nu)$ for the disjoint sets $\nu < \mu$ and $\nu \ge \mu$. By substituting the expressions of $P^1_{\mathcal{R}_1}(e|\mu, \nu)$ and $P^2_{\mathcal{R}_1}(e|\mu, \nu)$ provided in~\eqref{eq:marcumqnc} and $f_{\mu, \nu}(\mu, \nu)$ provided in~\eqref{Eq: jointdist}, we get the result given in the theorem.

Similarly, the average BER expression for receiver $\mathcal{R}_2$ is given by:
\begin{align}
P_{\mathcal{R}_2}(e) &= \int_{0}^{\infty} \int_{0}^{\infty} f_{\mu, \nu}(\mu, \nu) P_{\mathcal{R}_2}(e|\mu, \nu) \, \mathrm{d}\nu \, \mathrm{d}\mu.
\end{align}
Substituting the expressions of $P_{\mathcal{R}_2}(e|\mu, \nu)$ and $f_{\mu, \nu}(\mu, \nu)$ given in~\eqref{eq:marcumqdc} and~\eqref{Eq: jointdist} respectively, we get the result.
\end{IEEEproof} 

\section{Numerical Results and Discussion} \label{sec:NumResults}

\begin{figure}
    \centering
    \begin{subfigure}[b]{0.45\textwidth}
        \centering
        \includegraphics [width=\linewidth]{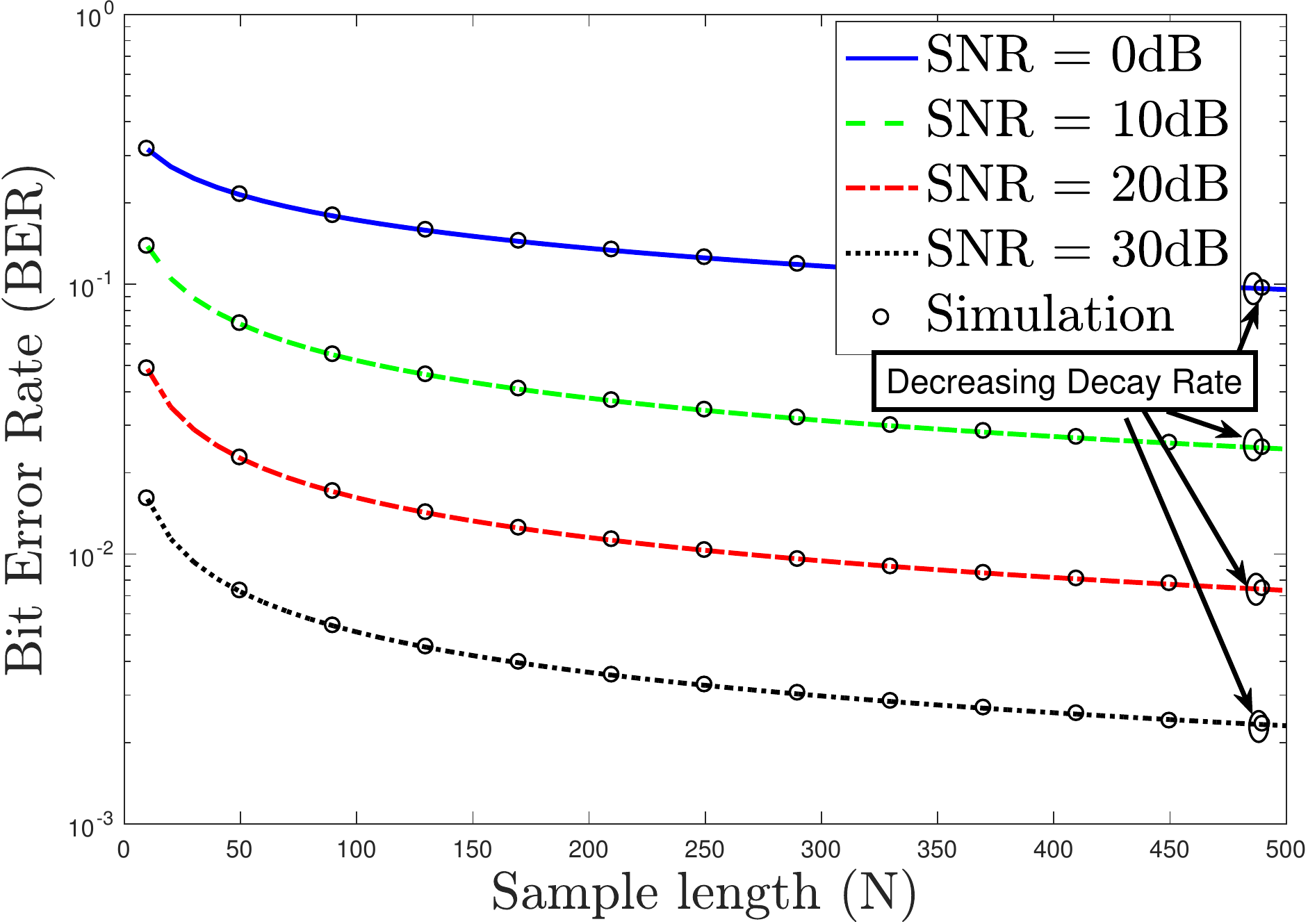}
         \caption{}\label{fig:ber_ncx2_mt_fading}
    \end{subfigure}
    ~ 
    \begin{subfigure}[b]{0.45\textwidth}
        \centering
        \includegraphics [width=\linewidth]{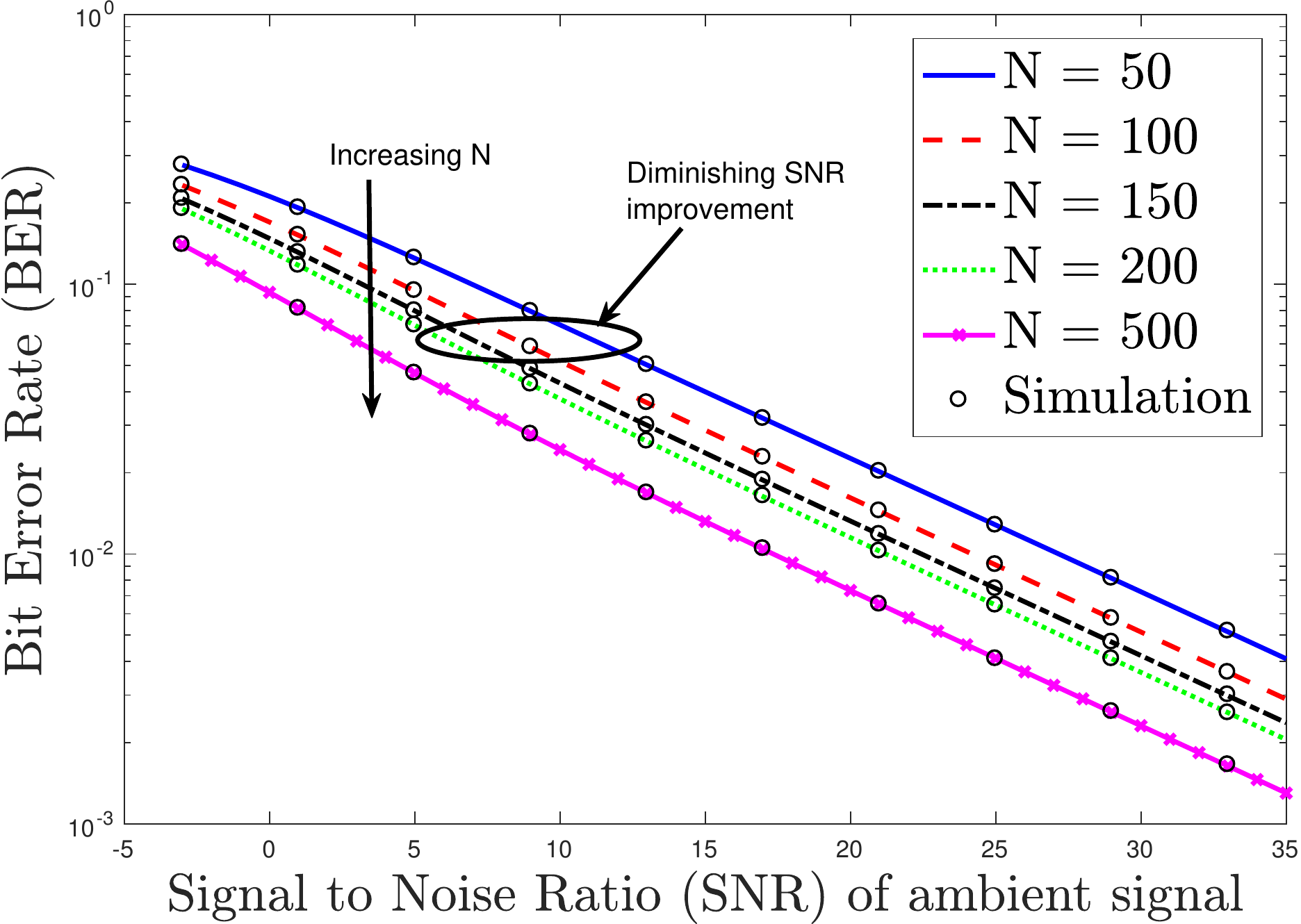}
	\caption{}\label{fig:ber_ncx2_mt_fading_snr}
    \end{subfigure}
    \caption{ Performance comparisons in MT technique: (a) BER versus $N$ for different SNR, (b) BER versus SNR for different $N$.}
\end{figure}
\begin{figure}
    \centering
    \begin{subfigure}[b]{0.45\textwidth}
        \centering
        \includegraphics [width=\linewidth]{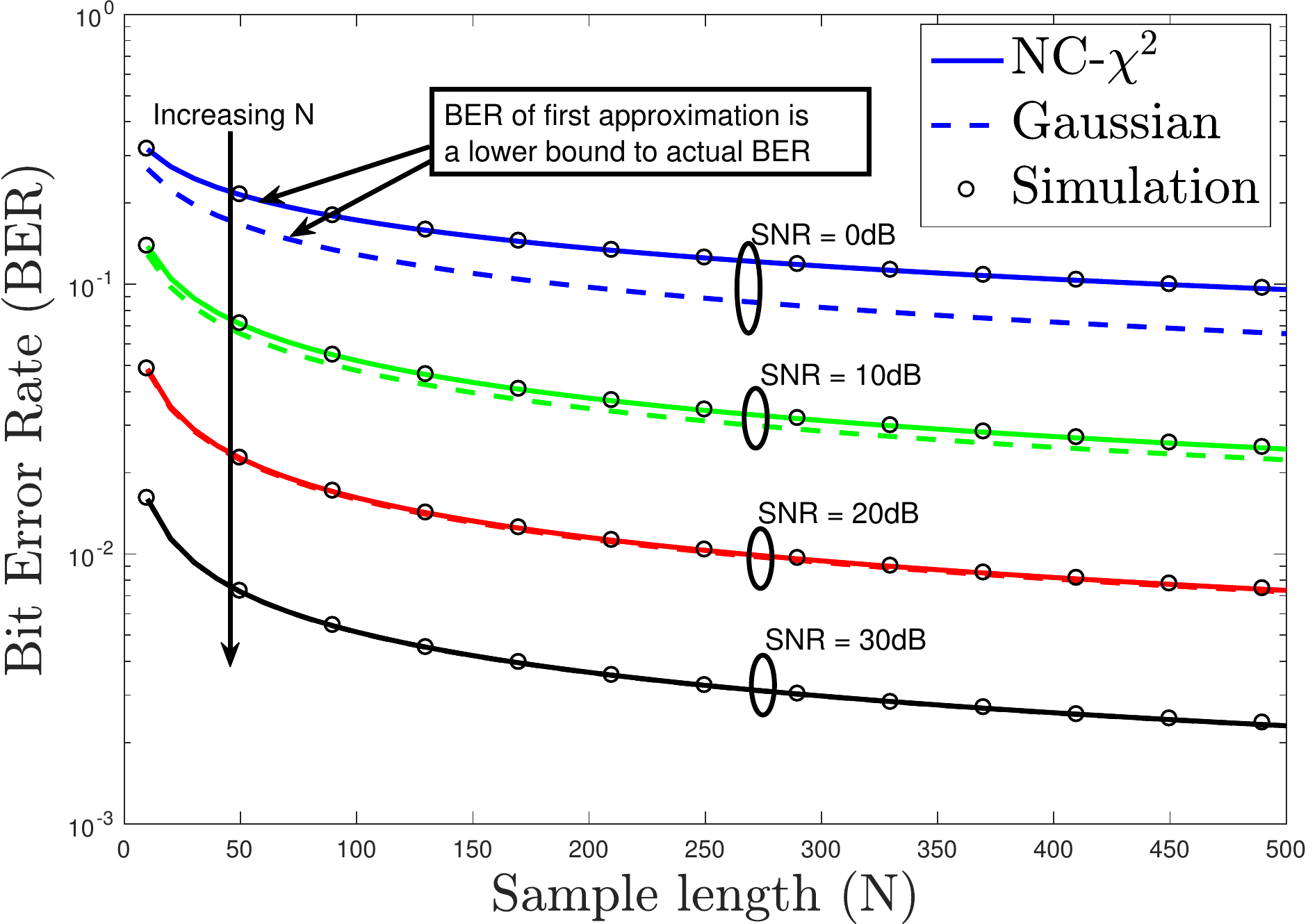}
         \caption{}\label{fig:ber_ncx2_ml_approx_comp_fading}
    \end{subfigure}
    ~ 
   \begin{subfigure}[b]{0.45\textwidth}
        \centering
        \includegraphics [width=\linewidth]{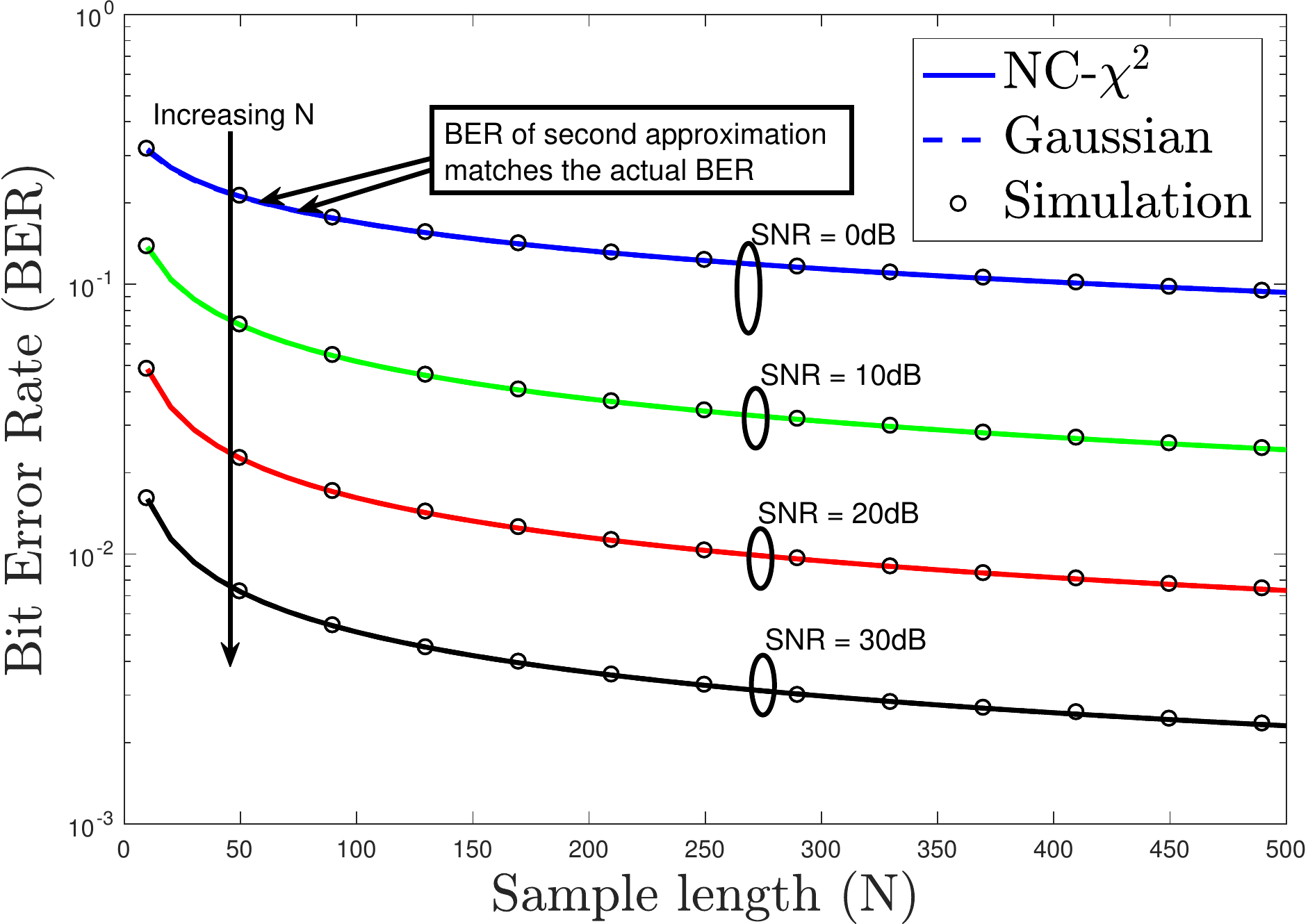}
         \caption{}\label{fig:ber_ncx2_ml_approx_comp_smallN_fading}
    \end{subfigure}
    \caption{ BER comparisons of actual and Gaussian approximated distributions for different SNR values using approximate ML threshold: (a) Actual vs first approximation, (b) Actual vs second approximation.}
\end{figure}
\begin{figure}
    \centering
    \begin{subfigure}[b]{0.45\textwidth}
        \centering
        \includegraphics [width=\linewidth]{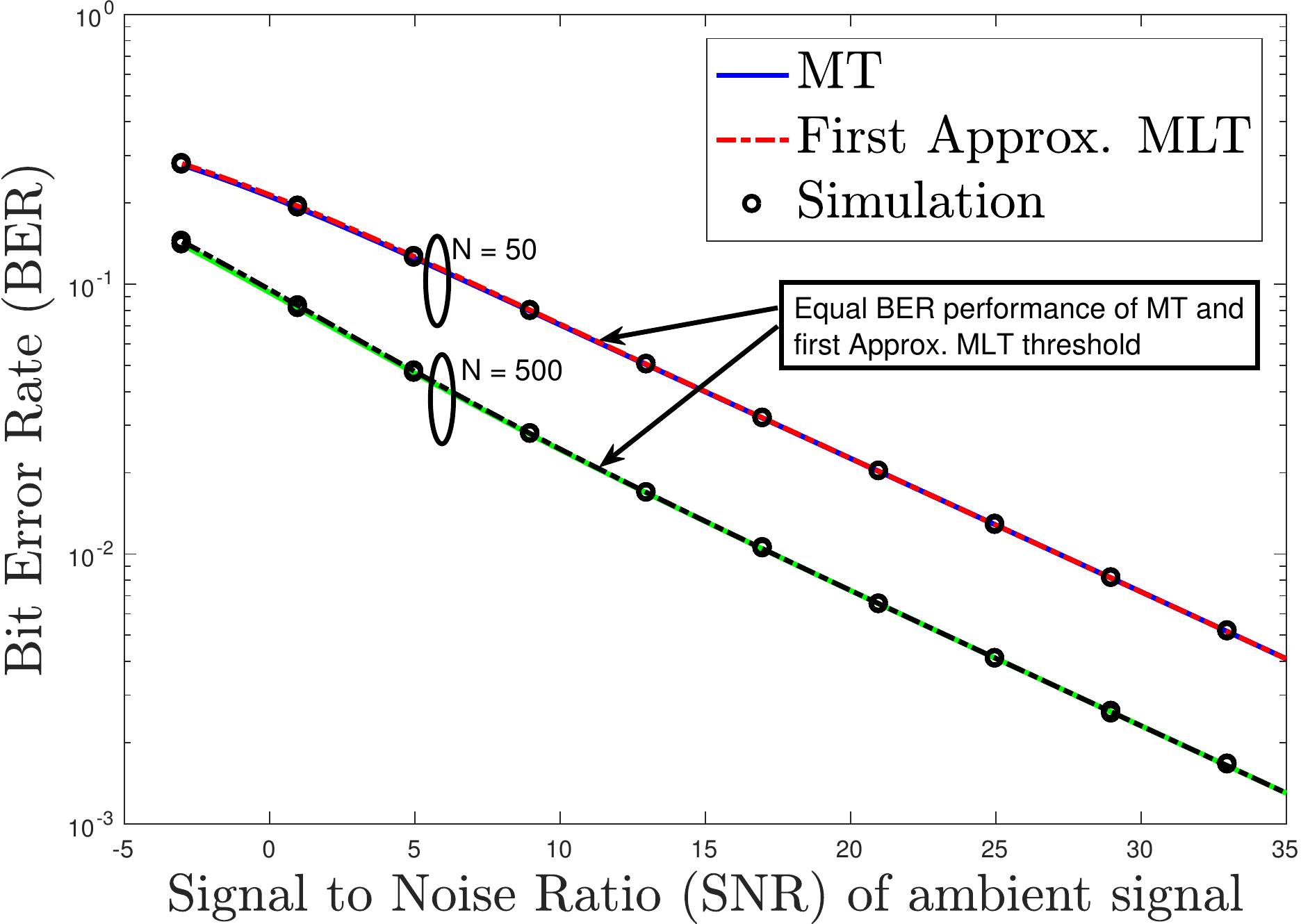}
         \caption{}\label{fig:ber_ncx2_mlapprox_mt_fading_snr}
    \end{subfigure}
    ~ 
    \begin{subfigure}[b]{0.45\textwidth}
        \centering
        \includegraphics [width=\linewidth]{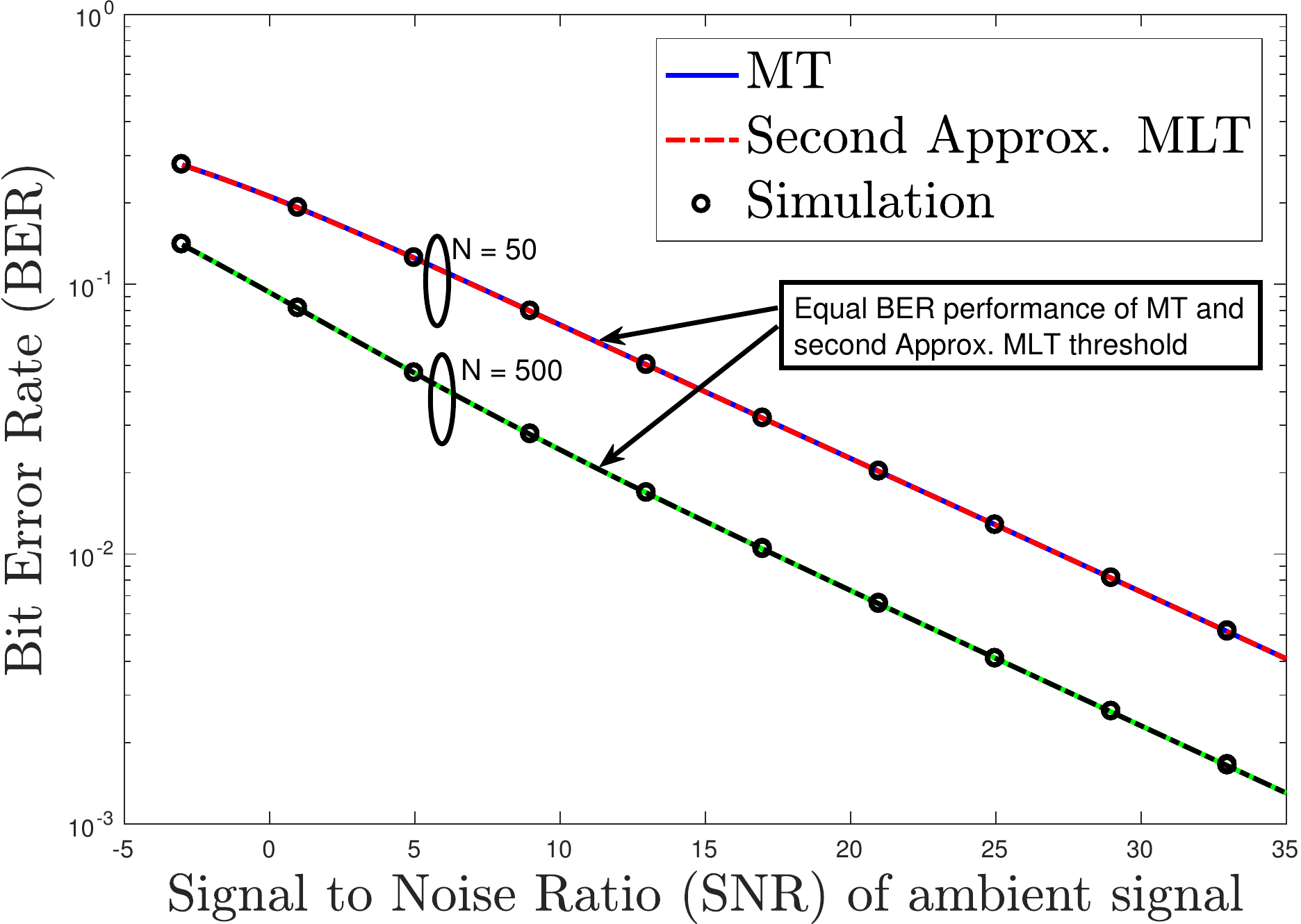}
	\caption{}\label{fig:ber_ncx2_mlapprox_mt_fading_snr_smallN}
    \end{subfigure}
    \caption{ Performance comparison of the two Approximate MLTs and MT at different values of $N$: (a) BER versus SNR for first approximate MLT and MT thresholds, (b) BER versus SNR for second approximate MLT and MT thresholds.}
\end{figure}
\begin{figure}
    \centering
    \begin{subfigure}[b]{0.45\textwidth}
        \centering
        \includegraphics [width=\linewidth]{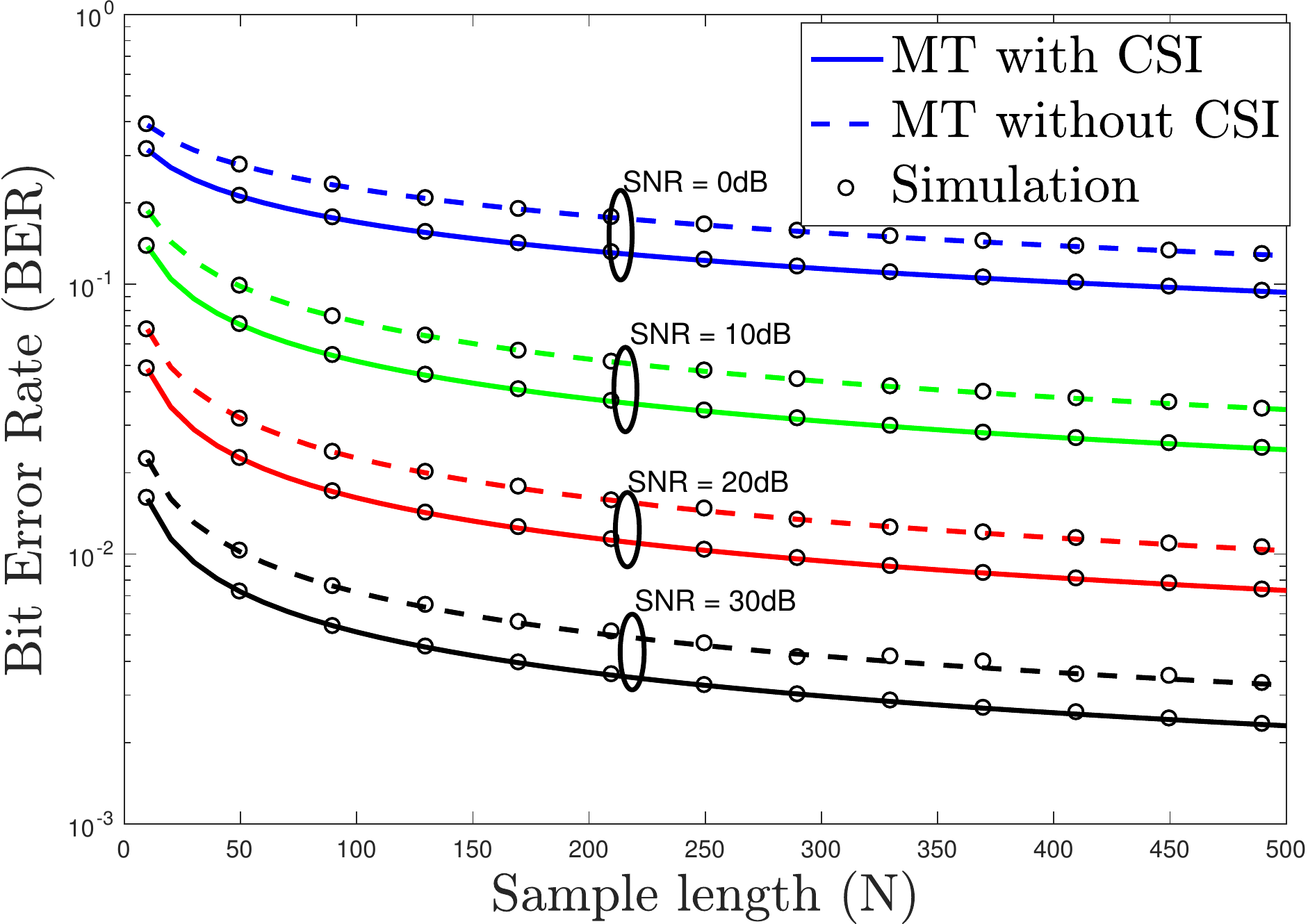}
         \caption{}\label{fig:ber_ncx2_mt_dpsk_fading}
    \end{subfigure}
    ~ 
    \begin{subfigure}[b]{0.45\textwidth}
        \centering
        \includegraphics [width=\linewidth]{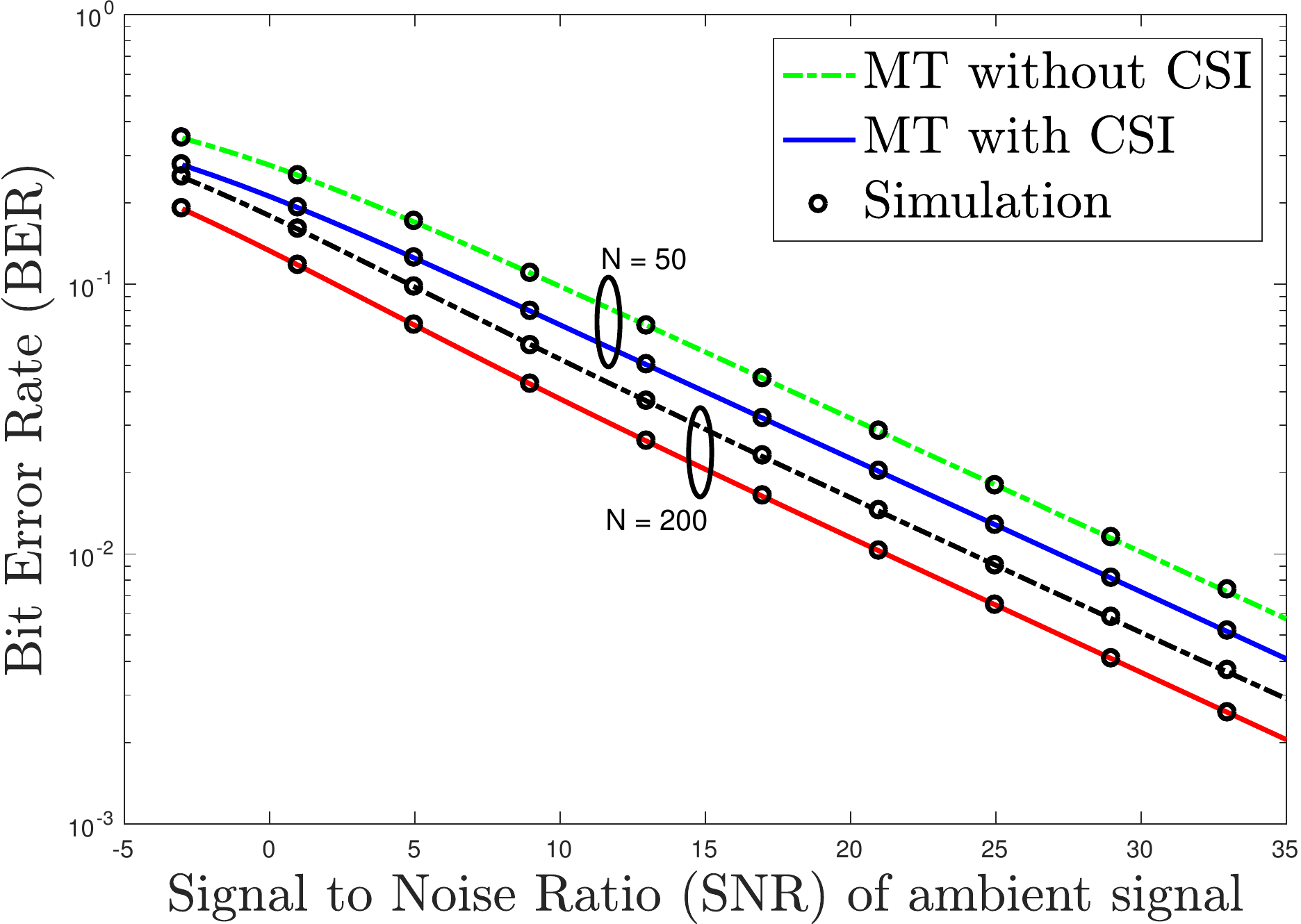}
	\caption{}\label{fig:ber_ncx2_mt_dpsk_fading_snr}
    \end{subfigure}
    \caption{Performance comparison of Receiver with CSI and Receiver without CSI in MT technique: (a) BER vs N, (b) BER vs SNR.}
\end{figure}

In this section, we plot the analytical results derived in the previous section to obtain useful system design insights. The analytical results are also validated by comparing with Monte Carlo simulations. The reflection coefficient $\alpha$ is set appropriately to approximate the 1.1 dB signal attenuation mentioned in \cite{shyam16} and the variance $\sigma_h^2$ of fading links is set to $1$ for the performance evaluation. First, the results of receiver $\mathcal{R}_1$ (with CSI) are presented before moving to receiver $\mathcal{R}_2$ (without CSI). With respect to any given system parameter, we refer to \emph{decay rate} as the rate of decrement in BER with the increasing value of that parameter. In Fig. \ref{fig:ber_ncx2_mt_fading}, we present the BER as a function of sample length $N$ for different SNR values. It can be observed that the decay rate decreases with respect to $N$. A similar comparison is shown in Fig. \ref{fig:ber_ncx2_mt_fading_snr} by plotting BER against SNR for different values of $N$. The gain in SNR of the system has diminishing returns with increasing $N$ as the performance of the energy averaging operation at the receiver converges to a limit, thereby limiting the improvement in BER. 

The difference in BER accuracy when using the approximated distributions instead of the exact distribution are compared in Figs. \ref{fig:ber_ncx2_ml_approx_comp_fading} and \ref{fig:ber_ncx2_ml_approx_comp_smallN_fading}. The first Gaussian approximation does not result in accurate BER at the lower SNR range as shown in Fig. \ref{fig:ber_ncx2_ml_approx_comp_fading}. The tightness of this approximation improves with increasing SNR. Further as shown in Fig. \ref{fig:ber_ncx2_ml_approx_comp_smallN_fading}, the second Gaussian approximation results in BER that is very accurate with respect to actual BER given by the exact distributions. For this reason, it can be concluded that the second Gaussian approximation should be the preferred mode of approximation out of the two at all values of $N$.

We now compare the BER performance of the threshold techniques MT and the two approximate MLTs. In particular, Figs. \ref{fig:ber_ncx2_mlapprox_mt_fading_snr} and \ref{fig:ber_ncx2_mlapprox_mt_fading_snr_smallN} depict the performance of the first approximate MLT and the second approximate MLT respectively compared to MT, from which we can conclude that both the approximate MLT techniques give similar BER performance as the MT technique. Hence, MT technique could be preferred due to the ease of implementation in either of the two receivers $\mathcal{R}_1$ and $\mathcal{R}_2$.

The performance of the two receivers $\mathcal{R}_1$ and $\mathcal{R}_2$ is compared in Figs. \ref{fig:ber_ncx2_mt_dpsk_fading} and \ref{fig:ber_ncx2_mt_dpsk_fading_snr}. As expected in the case of differential encoding, the performance of $\mathcal{R}_2$ is 3 dB worse than that of $\mathcal{R}_1$. The final insight from the analysis is that the BER of the optimal MT and other threshold techniques is dependent only on the received SNRs of the signal and not on the individual signal and noise energies. The technical discussion of this final insight is already presented in Remark \ref{rem:BERexp}.

\section{Conclusion}

In this paper, the error performance of an ambient backscatter system in a flat Rayleigh fading channel is characterized by deriving the exact analytical expressions of average BER both for the receivers with and without CSI. As part of the BER analysis, the exact conditional distributions of the average energy of the received signal is characterized in terms of the noncentral chi-squared distribution. The analysis requires careful treatment of the joint distribution of correlated fading components that appear in the two hypotheses in the BER derivation. Several key insights are drawn from the aforementioned analyses. First, the optimal BER of the ambient backscatter system is dependent on the energies of the signal and noise through SNR and not separately on the individual energies. Second, increasing the sample length $N$ provides diminishing returns in terms of BER improvement. 

This work has numerous extensions. First, the error analysis performed in this work is applicable only for slow varying channels. It is therefore important to extend it to fast fading scenarios as well. Second, in this work, we focused on the error performance of an isolated link. It is worthwhile to investigate if interference will have any noticeable impact on the BER in a dense IoT deployment. This analysis can perhaps be performed using tools from stochastic geometry. 
\appendix

\subsection{Proof of Lemma~\ref{lem:NCX2_1}} \label{app:NCX2_1}

The conditional PDF of $Y$ under $\mathcal{H}_0$ can be obtained from the conditional PDF of a scaled version given by $Z= \frac{Y}{c}$, where $c = \frac{\sigma^2}{2N}$. The expression of $Z$ can be written as follows:
\begin{align}
Z &= \frac{2}{\sigma^2} \sum_{n=1}^{N} |x(n) (h_r + h_b \alpha b h_t) + w(n)|^{2}.
\end{align}
Expanding $x(n) = x_r(n) + {\rm j} x_i(n)$, $h_0 = h_{0r}+ {\rm j} h_{0i}$ and $w(n) = w_r(n) + {\rm j} w_i(n)$, where ${\rm j} =\sqrt{-1}$, results in the form:
\begin{align}
	Z &= \frac{2}{\sigma^2} \sum_{n=1}^{N} |(x_r(n) + {\rm j} x_i(n))(h_{0r} + {\rm j} h_{0i} ) + w_r(n) + {\rm j} w_i(n)|^{2}, \nonumber\\
	&= \sum_{n=1}^{N} \frac{2}{\sigma^2} \left(x_r(n)h_{0r} - x_i(n) h_{0i} +w_r(n)\right)^2 + \sum_{n=1}^{N} \frac{2}{\sigma^2} \left(x_r(n)h_{0i} + x_i(n)h_{0r}  +w_i(n)\right)^2 \label{Eq: ncx},
\end{align}
where each term in the two summations is a square of an independent non-zero mean Gaussian RV with unit variance when conditioned on fading and $x(n)$. Also, notice that there are a total of $2N$ independent real-valued RVs.

The density function of this sum is given by noncentral chi-squared distribution \cite{ncx2wiki}. This distribution is associated with a non-centrality parameter $\lambda$ which is equal to the sum of the squared means of each Gaussian RV. The value of $\lambda$ corresponding to $Z$ can be evaluated as:
\begin{align}
	\lambda &= \dfrac{2\sum\limits_{n=1}^{N} \left(x_r(n)h_{0r} - x_i(n) h_{0i}\right)^2 }{\sigma^2} +  \dfrac{2\sum\limits_{n=1}^{N} \left(x_r(n)h_{0i} + x_i(n)h_{0r}\right)^2}{\sigma^2} \\
	& = \dfrac{2\sum\limits_{n=1}^{N} |x(n)|^2|h_0|^2}{\sigma^2} =  \dfrac{2\sum\limits_{n=1}^{N} |x(n)|^2 \mu}{\sigma^2} \stackrel{(a)}{=} \dfrac{2N \bar{E} \mu}{\sigma^2}.
\end{align}
where $(a)$ follows from the average energy given by~\eqref{eq: AvgE}. 

Notice that the distribution of $Z$ is independent of $x(n)$ since the parameter $\lambda$ approaches a constant value because of~\eqref{eq: AvgE}. Therefore, the PDF of $Z$ conditioned on $\mathcal{H}_0$ and $\mu$ is given by the noncentral chi-squared distribution with parameter $\lambda$ calculated above:
\begin{align}
	f_{Z|\mathcal{H}_0, \mu}(z) &= \sum \limits_{i=0}^{\infty} \dfrac{\exp(-\frac{\lambda}{2})(\frac{\lambda}{2})^{i}}{i!} f_{\chi^2}(z; 2N+2i) \nonumber\\
	&= \sum \limits_{i=0}^{\infty} \dfrac{\exp(-\frac{\mu N \bar{E} }{\sigma^2})(\frac{\mu N \bar{E}}{\sigma^2})^{i}}{i!} f_{\chi^2}(z; 2N+2i),
\end{align}
where $f_{\chi^2}(z; 2N+2i)$ is the PDF of central chi-squared distribution with degree $2N+2i$.

The conditional PDF $f_{Y|\mathcal{H}_0,\mu}(t)$ follows from the distribution of scaled transformation of a RV. The conditional PDF of $Y$ under $\mathcal{H}_1$ is derived using similar procedure. 

\subsection{Proof of Lemma~\ref{lem:MLthold}} \label{app:MLthold}
The distribution of a noncentral chi-square RV with degree $2v$ can be alternatively represented as a function of the modified Bessel function of the first kind $I_v(z)$ where $v$ represents order of the function. Hence, the conditional PDFs of average signal energy $Y$ whose distribution is characterized as noncentral chi-square with degree $2N$ can also be expressed as follows:
\begin{align}
f_{Y|\mathcal{H}_0, \mu}(t) &= \frac{N}{\sigma^2}  e^{-\left(\frac{N}{\sigma^2}t+\frac{N \mu \bar{E}}{\sigma^2}\right)} \left(\frac{4t}{\mu  \bar{E}}\right)^{\frac{N-1}{2}} I_{N-1}\left(\frac{2N}{\sigma^2} \sqrt{\mu  \bar{E} t}\right) \nonumber\\
&\stackrel{(e)}{=} \frac{N}{\pi \sigma^2}  e^{-\left(\frac{N}{\sigma^2}t+\frac{N \mu \bar{E}}{\sigma^2}\right)} \left(\frac{4t}{\mu  \bar{E}}\right)^{\frac{N-1}{2}} \int_{0}^{\pi} e^{\frac{2N}{\sigma^2} \sqrt{\mu \bar{E} t}\cos\theta} \cos (N-1)\theta \, \mathrm{d}\theta, \\
f_{Y|\mathcal{H}_1, \nu}(t) &= \frac{N}{\sigma^2}  e^{-\left(\frac{N}{\sigma^2}t+\frac{N \nu \bar{E}}{\sigma^2}\right)} \left(\frac{4t}{\nu  \bar{E}}\right)^{\frac{N-1}{2}} I_{N-1}\left(\frac{2N}{\sigma^2} \sqrt{\nu  \bar{E} t}\right) \nonumber\\
&\stackrel{(f)}{=} \frac{N}{\pi \sigma^2}  e^{-\left(\frac{N}{\sigma^2}t+\frac{N \nu \bar{E}}{\sigma^2}\right)} \left(\frac{4t}{\nu  \bar{E}}\right)^{\frac{N-1}{2}} \int_{0}^{\pi} e^{\frac{2N}{\sigma^2} \sqrt{\nu \bar{E} t}\cos\theta} \cos (N-1)\theta \, \mathrm{d}\theta,
\end{align}
where $(e)$ and $(f)$ follow from the integral form of the modified Bessel function of the first kind with integer order given for reference in definition \ref{def: FModBes1}.

By the ML rule, the threshold value $T_{\rm mlt}$ is chosen as the point where the two conditional distributions are equal and the simplified expression is given by the following equation:  
\begin{align}
e^{\frac{N}{\sigma^2}\bar{E}(\nu-\mu)} \left(\frac{\nu}{\mu}\right)^{\frac{N-1}{2}} \int_{0}^{\pi} e^{\frac{2N}{\sigma^2} \sqrt{\mu \bar{E} \, T_{\rm mlt}} \cos\theta} \cos (N-1)\theta\, \mathrm{d}\theta &= \int_{0}^{\pi} e^{\frac{2N}{\sigma^2} \sqrt{\nu \bar{E} \, T_{\rm mlt}} \cos\theta}  \cos (N-1)\theta\, \mathrm{d}\theta.
\end{align}

\subsection{Proof of Lemma~\ref{lem:APXMLthold}} \label{app:APXMLthold}
The approximations to the conditional PDFs  can be derived from~\eqref{eq:sumexp} which again is provided below for reference:
\begin{align*}
Y &= \underbrace{|h_r +  \alpha h_{b} h_t b|^2 \bar{E}}_{Y_0:\ \text{constant}}+\underbrace{\frac{2}{N} \operatorname{Re} \left\{\left(h_r +  \alpha h_{b} h_t b\right) \sum_{n=1}^{N} x(n) w^*(n) \right\}}_{Y_1:\ \text{Gaussian RV}} + \underbrace{\frac{1}{N} \sum_{n=1}^{N}  |w(n)|^2}_{Y_2 :\ \text{Central-$\chi^2$ RV }}.
\end{align*}
The conditional mean and variance of the Gaussian component $Y_1$ in the above equation is given by:
\begin{align}
\mathcal{H}_0 : \mathbb{E}[Y_1 |\mathcal{H}_0] = \mu \bar{E}, \text{VAR}[Y_1|\mathcal{H}_0] = \frac{2}{N} \mu \bar{E} \sigma^2,\\
	\mathcal{H}_1 :  \mathbb{E}[Y_1 |\mathcal{H}_1] = \nu \bar{E}, \text{VAR}[Y_1|\mathcal{H}_1] = \frac{2}{N} \nu \bar{E} \sigma^2.
\end{align}
The Central-$\chi^2$ component $Y_2$ will be approximated either as a constant or a Gaussian. In the first case (first Gaussian approximation), $Y_2$  can be simply approximated as the conditional mean of Central-$\chi^2$ RV which is $\sigma^2$. For the second case (second Gaussian approximation), $Y_2$ will be approximated as a Gaussian RV with conditional mean and variance equal to that of $Y_2$, as given below:
\begin{align}
\mathcal{H}_0 : \mathbb{E}[Y_2 |\mathcal{H}_0] = \sigma^2, \text{VAR}[Y_2|\mathcal{H}_0] = \frac{1}{N}\sigma^4,\\
	\mathcal{H}_1 :  \mathbb{E}[Y_2|\mathcal{H}_1] = \sigma^2, \text{VAR}[Y_2|\mathcal{H}_1] = \frac{1}{N}\sigma^4.
\end{align}
It is easy to see that $Y$ is Gaussian distributed under both approximations. For the first Gaussian approximation, the conditional distributions of $Y$ under the two hypotheses are given by:
\begin{align}
f_{Y|\mathcal{H}_0, \mu}(t) &= \frac{1}{\sqrt{2\pi\frac{2}{N}\mu \bar{E}\sigma^2}} \exp\left(-\frac{\left(t - \mu \bar{E} - \sigma^2\right)^2}{2 \frac{2}{N}\mu \bar{E}\sigma^2} \right), \label{eq:largeN1} \\
f_{Y|\mathcal{H}_1, \nu}(t) &= \frac{1}{\sqrt{2\pi\frac{2}{N}\nu \bar{E}\sigma^2}} \exp\left(-\frac{\left(t - \nu \bar{E} - \sigma^2\right)^2}{2 \frac{2}{N}\nu \bar{E}\sigma^2} \right). \label{eq:largeN2} 
\end{align}
Similarly, the conditional distributions of $Y$ under the two hypotheses for the second Gaussian approximation are given by:
\begin{align}
f_{Y|\mathcal{H}_0, \mu}(t) &= \frac{1}{\sqrt{2\pi \left(\frac{2}{N}\mu \bar{E}\sigma^2 + \frac{1}{N}\sigma^4 \right)}} \exp\left(-\frac{\left(t - \mu \bar{E} - \sigma^2\right)^2}{2  \left(\frac{2}{N}\mu \bar{E}\sigma^2 + \frac{1}{N}\sigma^4 \right)} \right), \label{eq:smallN1}\\
f_{Y|\mathcal{H}_1, \nu}(t) &= \frac{1}{\sqrt{2\pi \left(\frac{2}{N}\nu \bar{E}\sigma^2 + \frac{1}{N}\sigma^4 \right)}} \exp\left(-\frac{\left(t - \nu \bar{E} - \sigma^2\right)^2}{2  \left(\frac{2}{N}\nu \bar{E}\sigma^2 + \frac{1}{N}\sigma^4 \right)} \right). \label{eq:smallN2}
\end{align}
After equating the conditional distributions under the two hypotheses (separately for each of the approximations) and rearranging the terms, we get the final expressions of the threshold values.

\subsection{Proof of Lemma~\ref{lem:JointDist}} \label{app:JointDist}

We note that \cite{wang18} has derived the marginal distribution of $h_1$ and its magnitude squared parameter $\nu$ in the context of outage analysis for ambient backscatter systems. However, our derivation here is different since our focus is on the joint distribution of  $h_0$ and $h_1$, and their magnitude squared parameters $\mu$ and $\nu$ for the bit error rate analysis.

The distribution of independent and identical fading terms $h_r , h_t$ and $h_b$  is given by $\mathcal{CN} (0, \sigma_h^2)$. The distribution of $\alpha h_b \sim \mathcal{CN} (0,|\alpha|^2 \sigma_h^2)$, formed by combining $\alpha$ and $h_b$, follows from the scalar multiplication property of circularly symmetric Gaussian random vectors \cite[Sec. 7.8.1]{gallager08}.

The joint distribution of the real and imaginary parts of fading component $h_0$ is Gaussian. Similarly, the joint distribution of the real and imaginary parts of double Gaussian term $U = \alpha h_b  h_t$ of the fading component $h_1$ is given in \cite{pap65,moura12}. For completeness, the expressions are provided below:
\begin{align}
f_{h_{0R}, h_{0I}} (h_{0r}, h_{0i}) &= \dfrac{1}{\pi \sigma_h^2} \exp\left(-\dfrac{h_{0r}^2 + h_{0i}^2}{\sigma_h^2}\right), \\
f_{U_{R}, U_{I}} (u_{r}, u_{i}) &= \dfrac{1}{2\pi \left(\frac{|\alpha|\sigma_h^2}{2}\right)^2} K_0\left(\dfrac{\sqrt{u_{r}^2+ u_{i}^2}}{\frac{|\alpha|\sigma_h^2}{2}}\right),
\end{align}
where $K_0$ is the zeroth order modified Bessel function of second kind.

The joint distribution of the real and imaginary parts of $h_1$ conditioned on $h_0$ is related to the joint distribution of $U$ by the shift transformation property of a RV:
\begin{align}
	f_{h_{1R}, h_{1I}|h_{0R}, h_{0I}}(h_{1r}, h_{1i}) &= f_{U_{R}, U_{I}} ( h_{1r}- h_{0r} ,h_{1i}  - h_{0i}).
\end{align}
The joint distribution of the polar coordinates of $h_0$ and $h_1$ is derived from rectangular coordinates using the transformation property of RVs as follows:
\begin{align}
&f_{R_{h_0}, \Theta_{h_0}, R_{h_1}, \Theta_{h_1}} (r_{h_0}, \theta_{h_0}, r_{h_1}, \theta_{h_1})\nonumber \\
&\stackrel{(h)}{=} f_{R_{h_0}, \Theta_{h_0}} (r_{h_0}, \theta_{h_0}) \text{ } f_{R_{h_1}, \Theta_{h1}| R_{h_0}, \Theta_{h_0}} (r_{h_1}, \theta_{h_1}|r_{h_0}, \theta_{h_0})  \nonumber\\
&\stackrel{(i)}{=} r_{h_0} \text{ } f_{h_{0R}, h_{0I}} (r_{h_0} \cos \theta_{h_0}, r_{h_0} \sin \theta_{h_0}) \text{ } r_{h_1} \, f_{U_{R}, U_{I}} (r_{h_1} \cos \theta_{h_1} - r_{h_0} \cos \theta_{h_0}   r_{h_1} \sin \theta_{h_1} - r_{h_0} \sin \theta_{h_0})  \nonumber\\	
&= r_{h_0} \text{ }	\dfrac{1}{\pi \sigma_h^2} e^{-\frac{r_{h_0}^2}{\sigma_h^2}} \text{ } r_{h_1} \text{ } \dfrac{1}{2\pi \left(\frac{|\alpha|\sigma_h^2}{2}\right)^2} K_0\left(\dfrac{\sqrt{r_{h_1}^2 +  r_{h_0}^2 - 2 r_{h_1} r_{h_0} \cos (\theta_{h_1} - \theta_{h_0})}}{\frac{|\alpha|\sigma_h^2}{2}}\right),
\end{align}
where $(h)$ follows from de-conditioning of RVs through chain rule and $(i)$ follows from the relationship between the joint distribution functions of polar and rectangular coordinates.

The joint marginal distribution of $R_{h_1}, R_{h_0}$, obtained by integrating over the ranges of $\Theta_{h_0}$ and $\Theta_{h_1}$, is given by:
\begin{align}
&f_{R_{h_0}, R_{h_1}} (r_{h_0}, r_{h_1}) \nonumber \\
 &= \int_{0}^{2\pi} \int_{0}^{2\pi} \dfrac{r_{h_0}}{\pi \sigma_h^2} e^{-\frac{r_{h_0}^2}{\sigma_h^2}} \dfrac{r_{h_1}}{2\pi \left(\frac{|\alpha|\sigma_h^2}{2}\right)^2} K_0\left(\dfrac{\sqrt{r_{h_1}^2 +  r_{h_0}^2 - 2 r_{h_1} r_{h_0} \cos (\theta_{h_1} - \theta_{h_0})}}{\frac{|\alpha|\sigma_h^2}{2}}\right) \mathrm{d}\theta_{h_1} \mathrm{d}\theta_{h_0}.
\end{align}

Finally, the joint distribution of $\mu$ and $\nu$ is given by:
\begin{align}
f_{\mu, \nu}(\mu, \nu) &\stackrel{(j)}{=} \dfrac{1}{4\sqrt{\mu \nu}}f_{R_{h_0}, R_{h_1}} (\sqrt{\mu}, \sqrt{\nu})  \nonumber\\
&= \dfrac{1}{\pi \sigma_h^2} e^{-\frac{\mu}{\sigma_h^2}}  \dfrac{1}{2\pi (|\alpha|\sigma_h^2)^2} \int_{0}^{2\pi} \int_{0}^{2\pi} K_0\left(\dfrac{\sqrt{\mu +  \nu - 2 \sqrt{\mu \nu} \cos (\theta_{h_1} - \theta_{h_0})}}{\frac{|\alpha|\sigma_h^2}{2}}\right) \mathrm{d}\theta_{h_1} \mathrm{d}\theta_{h_0},
\end{align}
where $(j)$ follows from the relation between the joint PDFs of modulus of RVs given by $R_{h_0}$ and $R_{h_1}$, and the square of modulus of the same RVs given by $\mu$ and $\nu$, respectively.
\hfill 
\IEEEQED

\bibliographystyle{IEEEtran}
\bibliography{hokie}

\begin{thebibliography}{10}
\providecommand{\url}[1]{#1}
\csname url@samestyle\endcsname
\providecommand{\newblock}{\relax}
\providecommand{\bibinfo}[2]{#2}
\providecommand{\BIBentrySTDinterwordspacing}{\spaceskip=0pt\relax}
\providecommand{\BIBentryALTinterwordstretchfactor}{4}
\providecommand{\BIBentryALTinterwordspacing}{\spaceskip=\fontdimen2\font plus
\BIBentryALTinterwordstretchfactor\fontdimen3\font minus
  \fontdimen4\font\relax}
\providecommand{\BIBforeignlanguage}[2]{{%
\expandafter\ifx\csname l@#1\endcsname\relax
\typeout{** WARNING: IEEEtran.bst: No hyphenation pattern has been}%
\typeout{** loaded for the language `#1'. Using the pattern for}%
\typeout{** the default language instead.}%
\else
\language=\csname l@#1\endcsname
\fi
#2}}
\providecommand{\BIBdecl}{\relax}
\BIBdecl

\bibitem{shyam13}
V.~Liu, A.~Parks, V.~Talla, S.~Gollakota, D.~Wetherall, and J.~R. Smith,
  ``Ambient backscatter: Wireless communication out of thin air,'' \emph{Proc.,
  ACM SIGCOMM}, Aug. 2013.

\bibitem{shyam16}
B.~Kellogg, V.~Talla, S.~Gollakota, and J.~R. Smith, ``Ambient backscatter:
  Wireless communication out of thin air,'' \emph{Symposium on NSDI}, Mar.
  2016.

\bibitem{stock1948}
H.~Stockman, ``Communication by means of reflected power,'' \emph{Proc., of the
  IRE}, pp. 1196--1204, Oct. 1948.

\bibitem{gridur2009}
J.~D. Griffin and G.~D. Durgin, ``Complete link budgets for backscatter radio
  and {RFID} systems,'' \emph{IEEE Trans. on Antennas and Propagation
  Magazine}, vol.~51, no.~2, pp. 11--25, Apr. 2009.

\bibitem{smith2003}
D.~Kim, M.~A. Ingram, and W.~W.~S. Jr., ``Measurements of small-scale fading
  and path loss for long range {RF} tags,'' \emph{IEEE Trans. on Antennas and
  Propagation}, vol.~51, no.~8, pp. 1740--1749, Aug. 2003.

\bibitem{sahalos2014}
J.~Kimionis, A.~Bletsas, and J.~N. Sahalos, ``Increased range bistatic scatter
  radio,'' \emph{IEEE Trans. on Commun.}, vol.~62, no.~3, pp. 1091--1104, Mar.
  2014.

\bibitem{roym2014}
C.~Boyer and S.~Roy, ``Backscatter communication and {RFID}: coding, energy and
  {MIMO} analysis,'' \emph{IEEE Trans. on Commun.}, vol.~62, no.~3, pp.
  770--785, Mar. 2014.

\bibitem{gridur2008}
J.~D. Griffin and G.~D. Durgin, ``Gains for {RF} tags using multiple
  antennas,'' \emph{IEEE Trans. on Antennas and Propagation}, vol.~56, no.~2,
  pp. 563--570, Feb. 2008.

\bibitem{gridur2010}
------, ``Multipath fading measurement at 5.8 {GH}z for backscatter tags with
  multiple antennas,'' \emph{IEEE Trans. on Antennas and Propagation}, vol.~58,
  no.~11, pp. 3694--3700, Nov. 2010.

\bibitem{leigh2008}
G.~Vannucci, A.~Bletsas, and D.~Leigh, ``A software-defined radio system for
  backscatter sensor networks,'' \emph{IEEE Trans. on Wireless Commun.},
  vol.~2, no.~6, pp. 2170--2179, Jun. 2008.

\bibitem{peeters2014}
J.~Hermans, R.~Peeters, and B.~Preneel, ``Proper {RFID} privacy: model and
  protocols,'' \emph{IEEE Trans. Mobile Computing}, vol.~13, no.~12, pp.
  2888--2902, Dec. 2014.

\bibitem{saad2014}
W.~Saad, X.~Zhou, Z.~Han, and H.~V. Poor, ``On the physical layer security of
  backscatter wireless systems,'' \emph{IEEE Trans. on Wireless Commun.},
  vol.~13, no.~6, pp. 3442--3451, Jun. 2014.

\bibitem{shyam14}
A.~N. Parks, A.~Liu, S.~Gollakota, and J.~R. Smith, ``Turbocharging ambient
  backscatter communication,'' \emph{Proc., ACM SIGCOMM}, pp. 1--12, Aug. 2014.

\bibitem{shyam2014}
B.~Kellogg, A.~Parks, S.~Gollakota, J.~R. Smith, and D.~Wetherall, ``Wi-{F}i
  backscatter: Internet connectivity for {RF}-powered devices,'' \emph{Proc.,
  ACM SIGCOMM}, pp. 1--12, Aug. 2014.

\bibitem{liu14}
V.~Liu, V.~Talla, and S.~Gollakota, ``Enabling instantaneous feedback with
  full-duplex backscatter,'' \emph{Proc., ACM MobiCom}, no. 1-12, Sep. 2014.

\bibitem{shyam2016}
V.~Iyer, V.~Talla, B.~Kellogg, S.~Gollakota, and J.~R. Smith,
  ``Inter-technology backscatter: towards internet connectivity for implanted
  devices,'' \emph{Proc., ACM SIGCOMM}, pp. 1--14, Aug. 2016.

\bibitem{vamsi17}
V.~Talla, M.~Hessar, B.~Kellogg, A.~Najafi, J.~R. Smith, and S.~Gollakota,
  ``Lora backscatter: Enabling the vision of ubiquitous connectivity,''
  \emph{Proc., ACM on Interactive, Mobile, Wearable and Ubiquitous
  Technologies}, pp. 1--24, Sep. 2017.

\bibitem{varshney17}
A.~Varshney, O.~Harms, C.~Perez-Penichet, C.~Rohner, F.~Hermans, and T.~Voigt,
  ``Lorea: A backscatter architecture that achieves a long communication
  range,'' \emph{Proc., ACM on Embedded Network Sensor Systems (SenSys 17)},
  no.~50, Nov. 2017.

\bibitem{Wang15}
K.~Lu, G.~Wang, F.~Qu, and Z.~Zhong, ``Signal detection and {BER} analysis for
  {RF}-powered devices utilizing ambient backscatter,'' \emph{Proc., Intl.
  Conf. on Wireless Commun. \& Sig. Proc. (WCSP)}, Oct. 2015.

\bibitem{chintha15}
G.~Wang, F.~Gao, Z.~Dou, and C.~Tellambura, ``Uplink detection and {BER}
  analysis for ambient backscatter communication systems,'' \emph{Proc., IEEE
  Globecom}, Dec. 2015.

\bibitem{chintha16}
G.~Wang, F.~Gao, R.~Fan, and C.~Tellambura, ``Ambient backscatter communication
  systems: Detection and performance analysis,'' \emph{IEEE Trans. on Commun.},
  vol.~64, no.~11, pp. 4836 -- 4846, Nov. 2016.

\bibitem{gao16}
J.~Qian, F.~Gao, G.~Wang, S.~Jin, and H.~Zhu, ``Semi-{C}oherent {D}etection and
  {P}erformance {A}nalysis for {A}mbient {B}ackscatter {S}ystem,'' \emph{IEEE
  Trans. on Commun.}, vol.~65, no.~12, Dec. 2017.

\bibitem{gao17}
------, ``Noncoherent {D}etections for {A}mbient {B}ackscatter {S}ystem,''
  \emph{IEEE Trans. on Wireless Commun.}, vol.~16, no.~3, Mar. 2017.

\bibitem{hu2015}
Y.~Liu, Z.~Zhong, G.~Wang, and D.~Hu, ``Uplink detection and {BER} performance
  for wireless communication systems with ambient backscatter and multiple
  receiving antennas,'' \emph{Proc., Intl. Conf. on Commun. and Networking in
  China (ChinaCom)}, pp. 79 -- 84, Aug. 2015.

\bibitem{wang17}
Y.~Liu, G.~Wang, Z.~Dou, and Z.~Zhong, ``New {C}oding and {D}etection {S}chemes
  for {A}mbient {B}ackscatter {C}ommunication {S}ystems,'' \emph{IEEE Access},
  Mar. 2017.

\bibitem{chintha16vtc}
T.~Zeng, G.~Wang, Y.~Wang, Z.~Zhong, and C.~Tellambura, ``Statistical
  {C}ovariance {B}ased {S}ignal {D}etection for {A}mbient {B}ackscatter
  {C}ommunication {S}ystems,'' \emph{Proc., IEEE Veh. Technology Conf. (VTC)},
  Sep. 2016.

\bibitem{yang17}
G.~Yang, Y.-C. Liang, and Y.~P. Rui~Zhang, ``Modulation in the {A}ir:
  {B}ackscatter {C}ommunication over {A}mbient {OFDM} {C}arrier,'' \emph{IEEE
  Trans. on Commun.}, vol.~66, no.~3, Mar. 2018.

\bibitem{verde2016}
D.~Darsena, G.~Gelli, and F.~Verde, ``Modeling and {P}erformance {A}nalysis of
  {W}ireless {N}etworks {W}ith {A}mbient {B}ackscatter {D}evices,'' \emph{IEEE
  Trans. on Commun.}, vol.~65, no.~4, Apr. 2017.

\bibitem{kaibin17}
K.~Han and K.~Huang, ``Wirelessly powered backscatter communication networks:
  Modeling, coverage, and capacity,'' \emph{IEEE Trans. on Wireless Commun.},
  Mar. 2017.

\bibitem{seshu2005}
P.~V. Nikitin, K.~V.~S. Rao, S.~F. Lam, V.~Pillai, R.~Martinez, and
  H.~Heinrich, ``Power reflection coefficient analysis for complex impedances
  in {RFID} tag design,'' \emph{IEEE Trans. on Microw. Theory and Tech.},
  vol.~53, no.~9, Sep. 2005.

\bibitem{kurokawa1965}
K.~Kurokawa, ``Power waves and the scattering matrix,'' \emph{IEEE Trans. on
  Microw. Theory and Tech.}, vol. MTT-13, no.~3, Mar. 1965.

\bibitem{short12}
R.~T. Short, ``Computation of {R}ice and {N}oncentral {C}hi-{S}quared
  probabilities,'' \emph{PhaseLocked Systems, Technical Report PHS0254},
  vol.~16, no.~4, Apr. 2012.

\bibitem{shnidman89}
D.~A. Shnidman, ``The calculation of the probability of detection and the
  {G}eneralized {M}arcum {Q}-function,'' \emph{IEEE Trans. on Info. Theory},
  vol.~35, no.~2, Mar. 1989.

\bibitem{ncx2wiki}
M.~Abramowitz and I.~A. Stegun, ``Handbook of mathematical functions,''
  \emph{Appl. Math. Ser. 55, National Bureau of Standards}, pp. 942--943, Jun.
  1964.

\bibitem{wang18}
W.~Zhao, G.~Wang, S.~Atapattu, and C.~Tellambura, ``Outage analysis of ambient
  backscatter communication systems,'' \emph{Available online:
  arxiv.org/abs/1711.09214}.

\bibitem{gallager08}
R.~G. Gallager, ``Principles of digital communication,'' \emph{Cambridge
  University Press}, Jan. 2008.

\bibitem{pap65}
A.~Papoulis, ``Probability random variables and stochastic processes,''
  \emph{McGraw-Hill, 1st ed., p. 201}, 1965.

\bibitem{moura12}
N.~O'Donoughue and J.~M.~F. Moura, ``On the product of independent complex
  {G}aussians,'' \emph{IEEE Trans. on Signal Processing}, vol.~60, no.~3, Mar.
  2012.

\end{thebibliography}

\end{document}